\documentclass[fleqn,usenatbib]{mnras}

\usepackage{newtxtext,newtxmath}

\usepackage[T1]{fontenc}

\DeclareRobustCommand{\VAN}[3]{#2}
\let\VANthebibliography\thebibliography
\def\thebibliography{\DeclareRobustCommand{\VAN}[3]{##3}\VANthebibliography}

\usepackage{graphicx}	
\usepackage{amsmath}	
\usepackage{bm} 

\newcommand{\hi}{H\textsc{i}}
\newcommand{\cmfast}{\textsc{21cmfast}}

\title[EoR Power Spectrum Wedges]{Power spectrum multipoles and clustering wedges during the Epoch of Reionization}

\author[Z. Chen \& A. Pourtsidou]{
Zhaoting Chen $^{1}$\thanks{E-mail: zhaoting.chen@roe.ac.uk}
and Alkistis Pourtsidou $^{1,2}$
\\
$^{1}$Institute for Astronomy, The University of Edinburgh, Royal Observatory, Edinburgh EH9 3HJ, UK\\
$^{2}$Higgs Centre for Theoretical Physics, School of Physics and Astronomy, Edinburgh EH9 3FD, UK
}

\date{Accepted XXX. Received YYY; in original form ZZZ}

\pubyear{2024}

\begin{document}
\label{firstpage}
\pagerange{\pageref{firstpage}--\pageref{lastpage}}
\maketitle

\begin{abstract}
We study the viability of using power spectrum clustering wedges as summary statistics of 21\,cm surveys during the Epoch of Reionization (EoR).  {For observations in a wide redshift range $z\sim 7-9$ corresponding to a line-of-sight scale of $\sim 500\,$Mpc}, the power spectrum is subject to anisotropic effects due to the evolution along the light-of-sight. Information on the physics of reionization can be extracted from the anisotropy using the power spectrum multipoles. Signals of the power spectrum monopole are highly correlated at scales smaller than the typical ionization bubble, which can be disentangled by including higher-order multipoles. By simulating observations of  {the low frequency part of the Square Kilometre Array (SKA) Observatory}, we find that the sampling of the cylindrical wavenumber $\bm{k}$-space is highly non-uniform due to the baseline distribution,  {i.e. the distribution of antenna pairs sampling different transverse $\bm{k}_\perp$ scales}. Measurements in clustering wedges  {partition the cylindrical $\bm{k}$-space into different radial $k_\parallel$ scales, and} can be used for isolating parts of $\bm{k}$-space with relatively uniform sampling, allowing for more precise parameter inference. Using Fisher Matrix forecasts, we find that the reionization model can be inferred with per-cent level precision with $\sim 120$\,hrs of integration time using SKA-Low.  {Compared to model inference using only the power spectrum monopole above the foreground wedge}, model inference using multipole power spectra in clustering wedges yields a factor of $\sim 3$ improvement   {on the marginalised 1D} parameter constraints.
\end{abstract}




\section{Introduction}
 {The formation of luminous objects in the early Universe is key to the evolutionary history of the cosmic large scale structure.}
As the new generation of telescopes such as JWST \citep{2023PASP..135f8001G} reaches unprecedented sensitivity, for the first time a large number of very early galaxies at $8 \lesssim z \lesssim 17$ has been observed (e.g. \citealt{2022ApJ...938L..15C,2023ApJ...942L..27S,2023ApJ...946L..13F,2023ApJS..265....5H,2023MNRAS.518.4755A,2023ApJ...952L...7A}). 
 {The detection of a large population of galaxies hints towards tensions between the observations and the structure formation model in standard $\Lambda$ cold dark matter ($\Lambda$CDM) cosmology (e.g. \citealt{2024MNRAS.527.5929Y}).} 
However, resolving individual galaxies at very high redshifts requires observations with extreme depths, limiting the survey area of these observations and consequently the amount of information available for inferring cosmology.  {To compensate this fallback, complementary probes of the early Universe are needed to cover large volumes for cosmological studies of the early Universe.}

 {A promising candidate for probing large cosmological volumes in the early Universe is the cosmic 21\,cm line}. Neutral hydrogen (\hi) has an emission line at the rest wavelength of around 21\,cm \citep{1970ITIM...19..200H}. \hi\ is the most abundant element in the Universe after recombination (e.g. \citealt{1967ApJ...148....3W}), making the 21\,cm line an ideal tracer of the very early Universe (e.g. \citealt{2006PhRvL..97z1301C}). After the formation of the first luminous galaxies, \hi\ within the intergalactic medium (IGM) is ionized by the ultraviolet photons emitted by these objects during a period known as the Epoch of Reionization (EoR; see \citealt{2006PhR...433..181F} for a review). The redshifted 21\,cm line can be used to map the distribution of neutral hydrogen in the IGM, and therefore directly probes the EoR \citep{1997ApJ...486..581G}.

The 21\,cm line signal lies in the radio waveband and can be measured by radio telescopes.
However, it is intrinsically faint and requires very sensitive telescopes with large effective collecting area to detect it.
 {The requirements for large survey area and high angular resolution naturally call for using radio interferometers for 21\,cm observations \citep{1997ApJ...475..429M}.} Current experiments targeting the EoR include the Murchison Widefield Array \citep{2020MNRAS.493.4711T}, the Hydrogen Epoch of Reionization Array \citep{2023ApJ...945..124H} and the Low Frequency Array \citep{2020MNRAS.493.1662M}. In the future, the low frequency array of the SKA Observatory (SKAO; \citealt{2023JApA...44...43R}), the SKA-Low array, will have the sensitivity to produce high-fidelity tomographic image data for constraining the EoR \citep{2013ExA....36..235M}.
 {Measurements of the 21\,cm line signal are prone to observational effects.} Foreground emissions, i.e. continuum emissions from Galactic and extragalactic sources, are several orders of magnitude brighter than the \hi\ emission (see e.g. \citealt{2012MNRAS.423.2518C}). In order to mitigate the contamination from foregrounds, statistical techniques utilising the discrepancy in the spectral smoothness of foregrounds and the \hi\ signal are extensively studied. For interferometric observations, the technique of delay transform \citep{2004ApJ...615....7M}, which is Fourier transforming the measured visibility data along the frequency axis into delay space, is widely adopted (e.g. \citealt{2021MNRAS.505.4775Y,2023ApJ...945..124H}). 
The continuum foregrounds concentrate in low delay, leaving the high delay modes dominated by the \hi\ signal. Measuring the \hi\ signal by avoiding the low delay modes is referred to as foreground avoidance \citep{2012ApJ...752..137M}. On the other hand, blind signal separation techniques can be used to remove the smooth foregrounds, usually by quantifying the frequency-frequency covariance of the observed signal (e.g. \citealt{2020MNRAS.493.1662M,2022MNRAS.510.3495W,2023MNRAS.518.6262C}).

 {Both foreground avoidance and foreground removal techniques rely on the spectral smoothness of the foregrounds. Therefore,} systematic effects that break the spectral smoothness of foregrounds can severely contaminate 21\,cm observations. These effects include calibration errors (e.g. \citealt{2016MNRAS.461.3135B,2019ApJ...875...70B}), radio frequency interference (RFI; e.g. \citealt{2019PASP..131k4507W}), antenna cross-coupling \citep{2020ApJ...888...70K}, beam side-lobes \citep{2015ApJ...804...14T} and more.  {As a result, high-fidelity imaging of the IGM is prone to systematics, and current measurements of the EoR focus on obtaining the upper limits of the 21\,cm monopole power spectrum (see Figure 30 of \citealt{2023ApJ...945..124H} and references therein).}

 {As the detection of the 21\,cm signal closely intertwines with the mitigation of observational systematics, it is important to choose robust summary statistics of the 21\,cm signal to infer cosmology.} The spherically averaged power spectrum monopole is the most studied summary statistic for EoR, from theory (e.g. \citealt{2022MNRAS.513.5109G}), detection methodology (e.g. \citealt{2020PASP..132f2001L}) and model inference (e.g. \citealt{2015MNRAS.449.4246G}). However, the power spectrum monopole does not capture the full evolutionary information of the IGM. The growth of ionization bubbles in the IGM mimics a phase transition, where the ionized regions grow and become more interconnected with each other \citep{2016MNRAS.457.1813F,2019ApJ...885...23C}. The distribution of ionized regions is therefore highly non-Gaussian. Parameter inferences from the 1D power spectrum may exhibit biases and degeneracies between the astrophysical parameters \citep{2019MNRAS.484..933P}. Furthermore, the 21\,cm signal is observed along a 3D lightcone, and multiple observational effects introduce anisotropy into the 21\,cm signal such as redshift space distortions \citep{2012MNRAS.422..926M} and evolution along the line-of-sight \citep{2012MNRAS.424.1877D}. The anisotropy may bias parameter estimation \citep{2018MNRAS.477.3217G}, which cannot be fully quantified using the spherically averaged power spectrum.

 {In this work, we aim to provide alternative summary statistics to the spherically averaged monopole power spectrum for EoR.}
 {One common approach in the literature} is to use higher-order correlation functions for parameter inference, such as the bispectrum (e.g. \citealt{2017MNRAS.468.1542S,2022MNRAS.510.3838W}) and the trispectrum (e.g. \citealt{2008PhRvD..77j3506C}).  {The modelling of the covariance of higher-order correlation functions requires sophisticated simulations and good understanding of the underlying model (see e.g. \citealt{2024arXiv240109523P} for discussions in the context of galaxy surveys).} On the other hand, it has been suggested that tomographic images can be used for alternative summary statistics such as bubble size distributions (e.g. \citealt{2018MNRAS.473.2949G}). Topological analysis on the image data can also be used for fully extracting the underlying information, for example from Minkowski Functionals (e.g. \citealt{2017MNRAS.465..394Y}), shape finders (e.g. \citealt{2019MNRAS.485.2235B}) and Betti numbers \citep{2021MNRAS.505.1863G}. However, these summary statistics are more susceptible to observational effects. Even simple effects such as Gaussian thermal noise complicate the morphology of the tomographic image data (e.g. \citealt{2018MNRAS.479.5596G,2019ApJ...885...23C}). In general, topological quantities cannot be linearly formulated with regard to the data vector, making it difficult to reconstruct the signal and accurately model the covariance of the measurement.

 {In this paper, we argue that the power spectrum multipoles, measured in clustering wedges of the cylindrical $(\bm{k}_\perp,k_\parallel)$-space, can be used as a robust summary statistic for the EoR}.  {The power spectrum multipole is the 1D average of the cylindrical power spectrum, weighted by the Legendre polynomials of the cosine of the angle to the line-of-sight $\mu = k_\parallel/|\bm{k}|$.} It is the simplest extension to the 1D power spectrum monopole, and uses only the 2-point statistics of the 21\,cm signal.  {Therefore, techniques such as foreground avoidance and quadratic estimators can be applied to the multipoles in the same way as the monopole, making the power spectrum multipole immediately applicable to the data product of current and near-future surveys. }The power spectrum in 2D $(k_\perp,k_\parallel)$-space is found to be the most robust summary statistics for the 21\,cm signal \citep{2024arXiv240112277P}, which directly relates to the multipoles. As power spectrum multipoles measure the anisotropy of the 21\,cm signal \citep{2016MNRAS.456.2080M}, we expect that they may resolve the problem of parameter degeneracy, and improve the constraining power of 21\,cm observations. 

 {Furthermore, the $\bm{k}$-space for the weighted average of multipoles can be divided into multiple regions known as the clustering wedges (e.g. \citealt{2017MNRAS.467.2085G}), based on the values of $\mu$. By partitioning the $\bm{k}$-space into clustering wedges, the multipole power spectra can be used to disentangle the uneven sampling of $\bm{k}$-space, originated from the baseline distribution of radio interferometers.} We explore the viability of using future SKA-Low observations to measure the 21\,cm multipoles, and quantify the constraining power of the power spectrum clustering wedges on the EoR (see \citealt{Cunnington:2020mnn, Soares:2020zaq, Berti:2022ilk, 2023MNRAS.519.6246P} for post-reionization studies using the power spectrum multipoles).

The rest of this paper is organised as follows: 
In \hyperref[sec:sim]{Section \ref{sec:sim}}, we outline the procedure for simulating mock observations of SKA-Low.
The definitions of the power spectrum multipoles and clustering wedges are given in \hyperref[sec:psmulti]{Section \ref{sec:psmulti}}. 
We discuss the advantages of using the power spectrum multipoles and clustering wedges comparing to the spherically averaged monopole in \hyperref[sec:compare]{Section \ref{sec:compare}}. 
In \hyperref[sec:pars]{Section \ref{sec:pars}}, we present Fisher Matrix formalism for constraints on the EoR model parameters from multipole measurements. The results of the forecast are presented in \hyperref[sec:values]{Section \ref{sec:values}}. In \hyperref[sec:discussion]{Section \ref{sec:discussion}}, we further clarify and discuss the caveats of the simulations used. We summarise and conclude our findings in \hyperref[sec:conclusion]{Section \ref{sec:conclusion}}. 
Throughout this paper, we assume the flat $\Lambda$CDM cosmology reported in \cite{2020A&A...641A...6P}.

\section{Simulation}
\label{sec:sim}
\subsection{The 21\,cm signal}
\label{subsec:21cmsim}
In this section, we describe the simulation of the 21\,cm signal used in this paper. The 21\,cm signal during the Epoch of Reionization is represented by the brightness temperature fluctuation against the background cosmic microwave background (CMB) temperature
\begin{equation}
    \delta T_b = \frac{T_{S}-T_\gamma}{1+z}\big(1-{\rm exp}[-\tau_{\mu_0}]\big),
\end{equation}
where $T_{S}$ is the gas spin temperature, $T_\gamma$ is the CMB temperature, $z$ is the redshift of the signal, and $\tau_{\mu_0}$ is the optical depth at the rest wavelength of the 21\,cm signal $\mu_0$. In the optically thin approximation $\tau_{\mu_0}\ll 1$, the brightness temperature can be written as (e.g. \citealt{2006PhR...433..181F})
\begin{equation}
\begin{split}
    \delta T_b \approx & 27{\rm mK}\,x_{\rm \hi}(1+\delta_m)\bigg(1-\frac{T_\gamma}{T_S}\bigg)\bigg( \frac{1+z}{10}\frac{\Omega_m}{0.27} \bigg)^{\frac{1}{2}}\\
    & \times \bigg( \frac{\Omega_b h}{0.44\times 0.7} \bigg) \bigg(1+\frac{{\rm d}v_\parallel/{\rm d}r}{H}\bigg)^{-1},
\end{split}
\label{eq:tb}
\end{equation}
where $x_{\rm \hi}$ is the neutral fraction of the IGM, $\delta_m$ is the matter overdensity, $\Omega_m$ is the fraction of matter density in total energy density at present day, $\Omega_b$ is the fraction of baryon density in total energy density at present day, $H$ is the Hubble expansion rate, $h = H_0 /(100\,{\rm km/s/Mpc})$ where $H_0$ is the Hubble parameter at $z=0$, and ${\rm d}v_\parallel/{\rm d}r$ is the gradient of the peculiar velocity field along the line-of-sight.

We use the publicly available software \cmfast\footnote{\url{https://21cmfast.readthedocs.io}} \citep{2011MNRAS.411..955M,2020JOSS....5.2582M} to simulate the 21\,cm signal. \cmfast\ uses the excursion set formalism \citep{2004ApJ...613....1F} to enable fast, semi-numerical computations of the reionization process. In particular, it computes the neutral fraction of the IGM by iteratively checking for the maximum radius $R$ around a cell $\bm{x}$ where the criterion (e.g. \citealt{2015MNRAS.449.4246G})
\begin{equation}
    \zeta \, f_{\rm coll}(\bm{x},z,R,\bar{M}_{\rm min}) > 1
\end{equation}
can be met. Here, $\zeta$ is the ionization efficiency parameter to be discussed later, and $f_{\rm coll}$ is the collapse function describing the fraction of collapsed matter within a spherical radius R residing in halos with mass larger than $\bar{M}_{\rm min}$ (e.g. \citealt{1974ApJ...187..425P, 1999MNRAS.308..119S}). If the above criterion is met, the cell is fully ionized. If not, the ionized fraction of the cell is set to be $\zeta \, f_{\rm coll}(\bm{x},z,R_{\rm cell},\bar{M}_{\rm min})$.

In our simulation settings, the collapse function is tuned by the minimum virial temperature of the star-forming halos, $T_{\rm vir}$ (see e.g. \citealt{2001PhR...349..125B}). Higher $T_{\rm vir}$ results in fewer halos to produce ionizing photons, requiring a larger ionization efficiency $\zeta$ to ionize the IGM.

The ionization efficiency parameter $\zeta$ is an accumulation of multiple physical properties of the ionizing sources
\begin{equation}
    \zeta = 30\frac{f_{\rm esc}}{0.3}\frac{f_*}{0.05}\frac{N_\gamma}{4000}\frac{2}{1+\bar{n}_{\rm rec}(\bm{x},z,R)} = \xi /(1+\bar{n}_{\rm rec}(\bm{x},z,R)),
\end{equation}
where $f_{\rm esc}$ is the fraction of ionizing photons that escape into the IGM, $f_*$ is the fraction of galactic gas in stars, $N_\gamma$ is the number of ionizing photons produced per baryon in stars, and $\bar{n}_{\rm rec}(\bm{x},z,R)$ is the average number of times an \hi\ atom recombines. In our simulation settings, we allow inhomogeneous recombination and calculate the recombination number following the procedures in \cite{2014MNRAS.440.1662S}, while keeping the rest of the ionization efficiency $\xi$ fixed.

As we focus on power spectrum multipoles, the anisotropic effects need to be included in the simulation accurately. The evolution along the observational lightcone produces sizeable anisotropy (e.g. \citealt{2012MNRAS.424.1877D,2014MNRAS.442.1491D}). Furthermore, redshift space distortions due to the peculiar velocity field also have a major impact on the 21\,cm signal \citep{2012MNRAS.422..926M,2013MNRAS.434.1978M}. Therefore, we follow \cite{2013MNRAS.435..460J} and simulate the full brightness temperature fluctuation without the optically thin approximation for a 3D lightcone. The simulation voxel is split into sub-voxels to apply redshift space distortion effects according to the simulated velocity field, and the perturbed sub-voxels are regridded to the simulation box.

Throughout this paper, the simulations have coeval boxes with $(1600\,{\rm Mpc})^3$ size and $(2\,{\rm Mpc})^3$ resolution. The $1600\,{\rm Mpc}$ size corresponds to $\sim 10\,{\rm deg}$ for $z\sim 8$, which is at $\sim 3\sigma$ of the power beam of SKA-Low discussed later in \hyperref[subsec:obssim]{Section \ref{subsec:obssim}}. The $2\,{\rm Mpc}$ resolution ensures that the multipoles can be accurately modelled down to $k\sim 1\,{\rm Mpc^{-1}}$.

\begin{figure*}
    \centering
    \includegraphics[width=\linewidth]{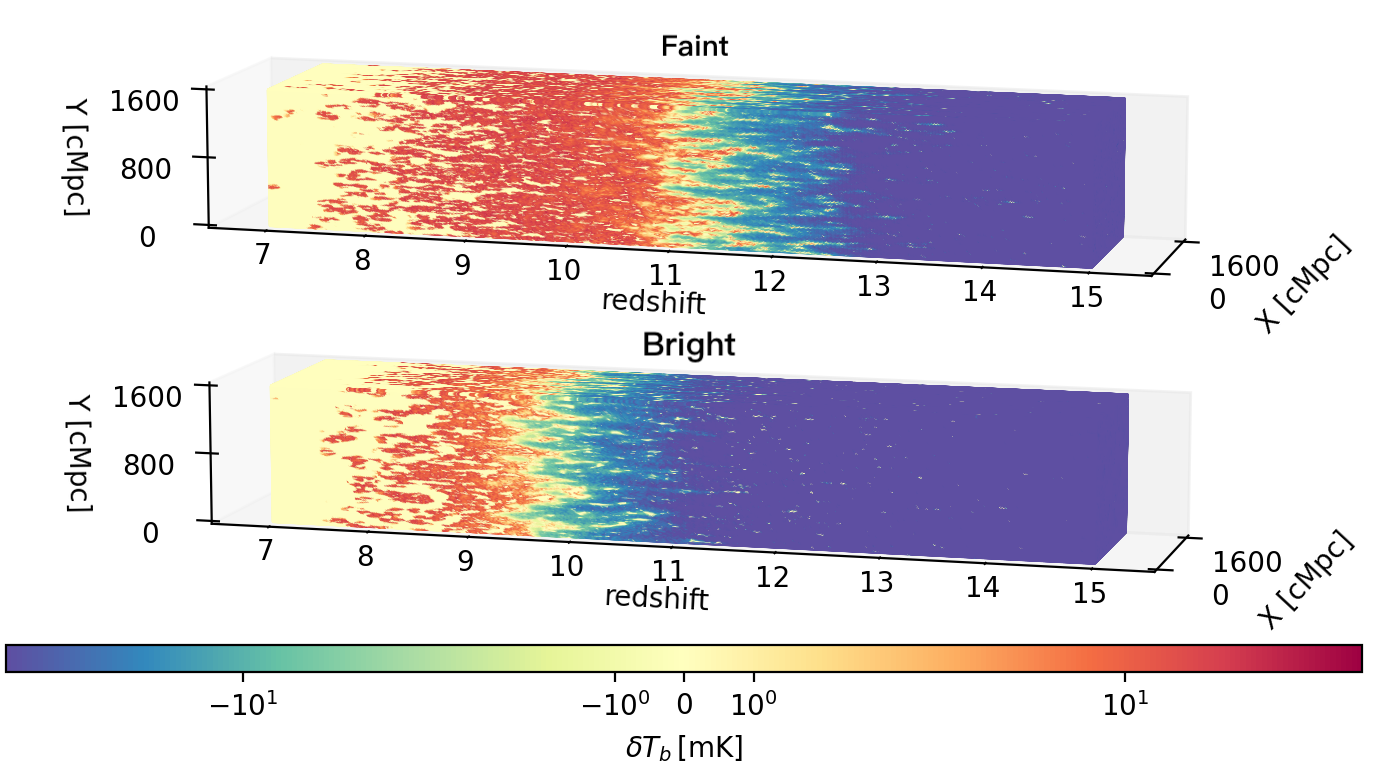}
    \caption{ {A visualization of the evolution of the 21\,cm brightness temperature $\delta T_b$ for our simulations with $({\rm log[T_{vir}]}, \xi) = (4.7,65)$ (faint) and for $({\rm log[T_{vir}]}, \xi) = (5.1,150)$ (bright) for one realization. The maximum and minimum values of the colour bar are set to $\pm 50 \,$mK for better visualization. Note the symmetric logarithmic scale for the temperature field.}}
    \label{fig:lightcone}
\end{figure*}

\begin{figure}
    \centering
    \includegraphics[width=\linewidth]{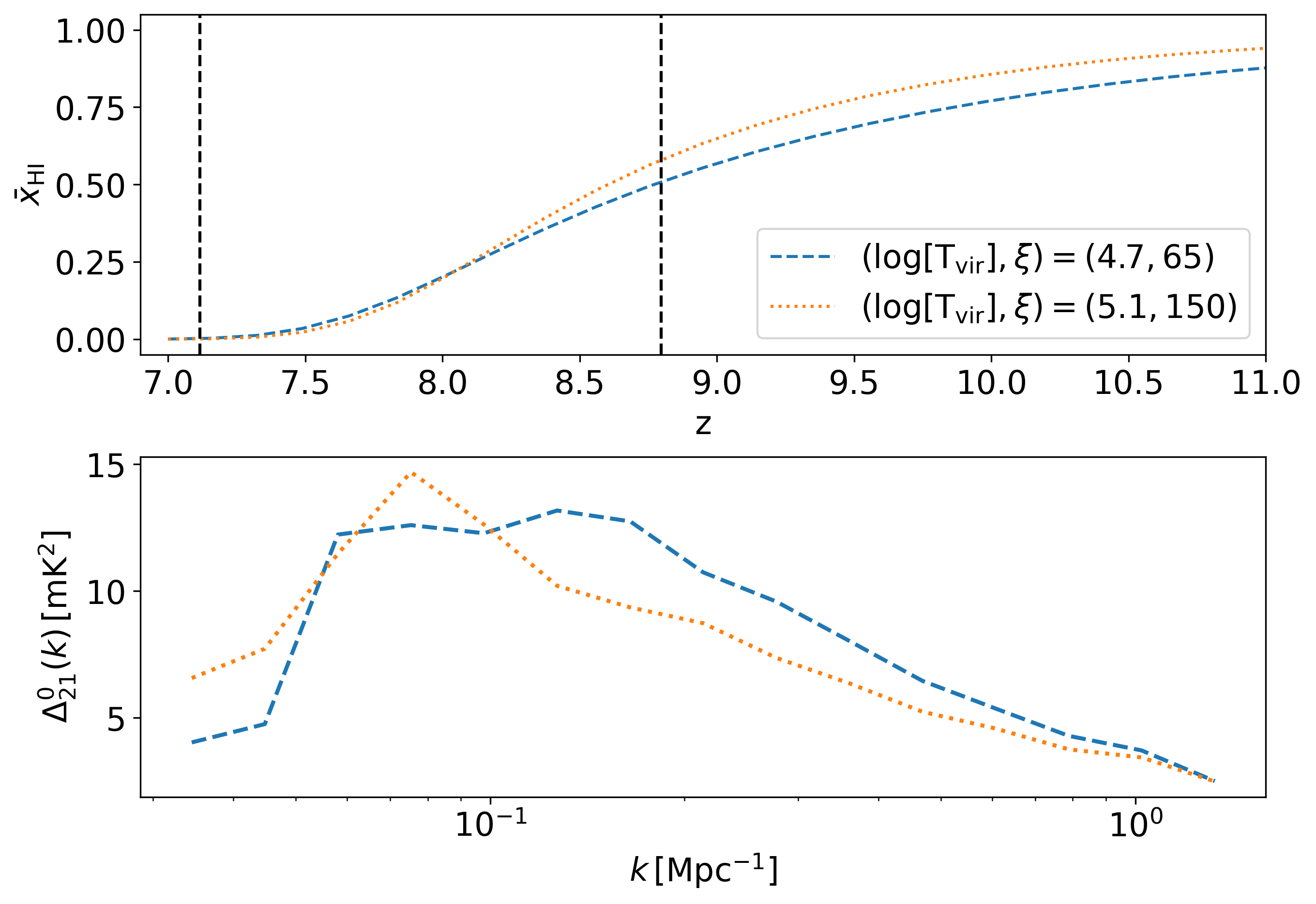}
    \caption{Top panel: The global ionization fraction of the IGM for $({\rm log[T_{vir}]}, \xi) = (4.7,65)$ (blue dashed line) and for $({\rm log[T_{vir}]}, \xi) = (5.1,150)$ (yellow dotted line). The black vertical dashed lines denote the $z=7.1$ and $z=8.8$ boundaries, which we use for the visibility simulation. Bottom panel: The spherically averaged 1D power spectrum of the brightness temperature field defined in \autoref{eq:psdimless} for the $z=7.1-8.8$ lightcone.}
    \label{fig:glob}
\end{figure}

An important aspect of our work is to illustrate the constraining power of the power spectrum multipoles for EoR parameters. Therefore, we choose two different sets of EoR parameters $({\rm log_{10}[T_{vir}/K]}, \xi)$, which produce an almost identical global ionization history as well as a similar spherically averaged 1D power spectrum for comparison. Specifically, we choose $({\rm log_{10}[T_{vir}/K]}, \xi) = (4.7,65),\,(5.1,150)$ to produce two sets of reionization lightcones.
This is similar to the ``faint galaxies'' and ``bright galaxies'' models presented in \cite{2018MNRAS.477.3217G}. For simplicity, we use the same names for the two models we are considering. Lower values of $({\rm log_{10}[T_{vir}/K]}, \xi)$ mean that the reionization process is driven by fainter, more numerous sources, and vice versa. An illustration of the differences in the reionization process between the two models is shown in \autoref{fig:lightcone}.  {As seen in \autoref{fig:lightcone}, the brightness temperature field indeed has distinct morphologies between the faint and bright models, where the reionization process in the faint model is more homogeneous.} The evolution of the global ionization fraction and the 1D monopole power spectrum along the observation lightcone are presented in \autoref{fig:glob} for reference\footnote{For simplicity, from here on we use ${\rm log [T_{vir}]}$ to denote ${\rm log_{10}[T_{vir}/K]}$.}.

We note that, in our simulation, reionization is almost complete at $z\sim 7.0$, which is faster than what has been suggested by current measurements (e.g. \citealt{2022MNRAS.514...55B,2024arXiv240412585G}). As discussed later in \hyperref[subsec:obssim]{Section \ref{subsec:obssim}}, we focus on the redshift range of $z\sim 7.1-8.8$ and thus a faster reionization produces larger anisotropic effects along the lightcone, which is ideal for presenting the case of comparing the constraining power of the monopole and the multipoles.

\subsection{The interferometric observation}
\label{subsec:obssim}
In this section, we present the simulation of visibility data using the configurations of the SKA-Low array. The SKA-Low array will be located at the Murchison Radio-astronomy Observatory\footnote{\url{https://www.skao.int/en/explore/telescopes/ska-low}} (MRO), sharing the location with its precursor, the Murchison Widefield Array (MWA; \citealt{2013PASA...30....7T}). SKA-Low is capable of measuring the 21\,cm signal from 50\,MHz to 350\,MHz, covering $z\sim 3-25$. The radio environment at the MRO provides a relatively clean frequency range of 140-200\,MHz for EoR measurements \citep{2015PASA...32....8O,7386856}. We choose the sub-band of 145-175\,MHz, corresponding to the redshift range of $z\sim 7.1-8.8$ towards the later stage of the EoR when more than half of the IGM is ionized \citep{2018ApJ...864..142D}. The channel bandwidth is to set to 200\,kHz, corresponding to the line-of-sight resolution of $k_\parallel\lesssim 1\,{\rm Mpc}^{-1}$. While the frequency resolution of SKA-Low will be finer than 200\,kHz, the nonlinear scales $k_\parallel > 1\,{\rm Mpc}^{-1}$ are hard to model and beyond the scope of this work.
The 21\,cm lightcone is smoothed along the line-of-sight to match the frequency resolution.

\begin{figure}
    \centering
    \includegraphics[width=\linewidth]{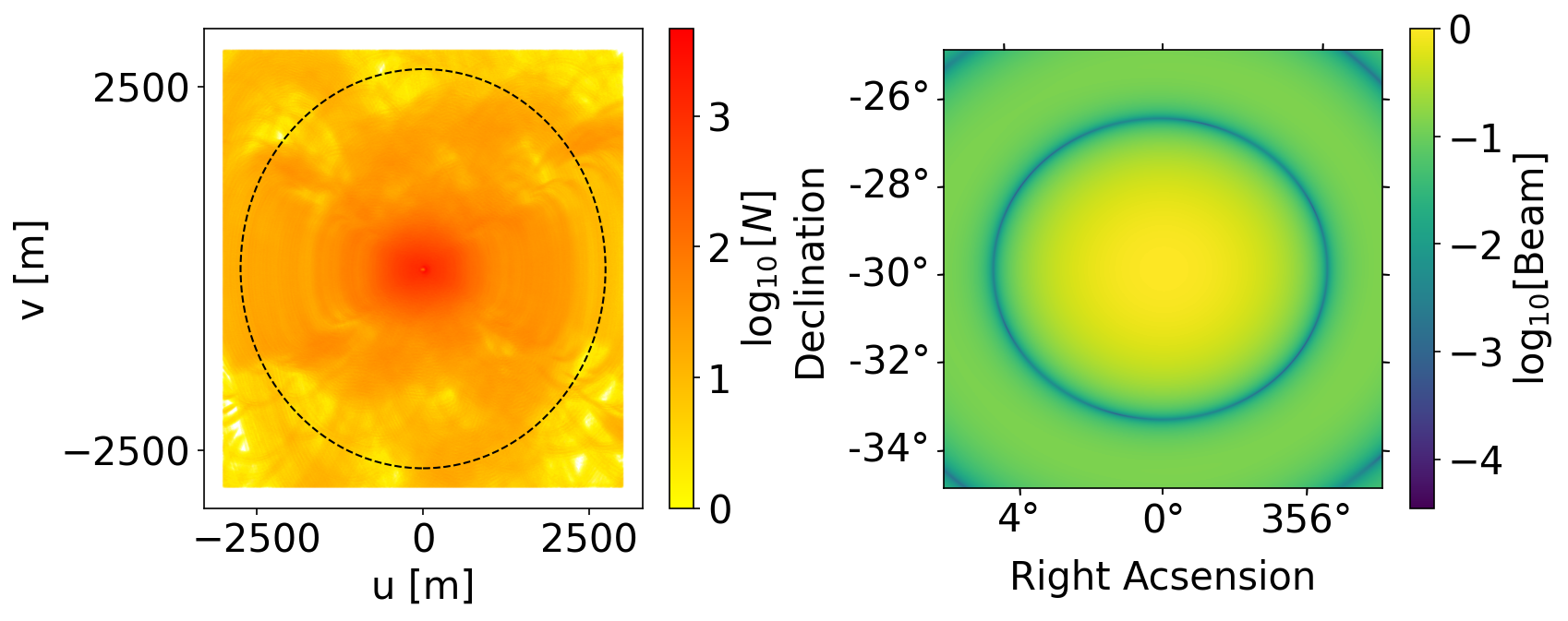}
    \caption{Left panel: The baseline distribution of the simulated tracking observation for SKA-Low. The grid size is chosen to be $(6\,{\rm m})^2$ for illustration, and only the central area of $(3000\,{\rm m})^2$ is shown. The dashed circle indicates the $k_\perp \sim 1\,{\rm Mpc^{-1}}$ boundary within which the power spectra are calculated. Right panel: The instrument power beam of SKA-Low around the pointing centre.}
    \label{fig:beam}
\end{figure}

We use the \textsc{oskar} package\footnote{\url{https://github.com/OxfordSKA/OSKAR}} \citep{5613289} to simulate the interferometric observations, with the telescope configuration following the latest official SKAO data challenge\footnote{\url{https://sdc3.skao.int/challenges/foregrounds/data}}. The pointing centre is set to be near the EoR0 field \citep{2021PASA...38...57L} at (RA=0\,deg, Dec=-30\,deg). The $u$-$v$ coverage of the baseline distribution is simulated with a tracking observation of 12\,hrs with a time resolution of 10\,s. The baseline distribution and the power beam are shown in \autoref{fig:beam} for reference.

The standard deviation of the Gaussian thermal noise of visibilities follows the radiometer equation (e.g. \citealt{9d4a7bd4-8744-3e49-ae1c-0eaf2bd342e7})
\begin{equation}
    \sigma_{\rm N} = \frac{2k_B T_{\rm sys}}{A_{\rm e}\sqrt{\delta t \delta f}},
\label{eq:sigman}
\end{equation}
where $T_{\rm sys}$ is the system temperature, $A_{\rm e}$ is the effective collecting area of the antenna, $\delta t = 10\,$s is the time resolution and $\delta f = 200$\,kHz is the channel bandwidth. In this paper, we aim to provide forecasts for power spectrum multipole measurements of a deep observation with coherent averaging of multiple nights. For a total integration time of $t_{\rm tot}$, the thermal noise amplitude for each baseline is scaled so that
\begin{equation}
    \sigma_{\rm N}^{\rm avg} = \sigma_{\rm N}\sqrt{\frac{t_{\rm n}}{t_{\rm tot}}},
\end{equation}
where $t_{\rm n} = 12\,$hrs is the observation time of one night. In this paper, we follow the anticipated performance of SKA-Low in \cite{2019arXiv191212699B} to produce the thermal noise. The natural sensitivity of the instrument is set to be $A_{\rm e}/T_{\rm sys} = [1.052,1.119,1.159,1.171,1.190]\,{\rm m^2/K}$ at $[140,150,160,170,180]$\,MHz, and the amplitudes at each frequency channels are linearly interpolated. The total observation time is chosen to be $t_{\rm tot} = 120\,$hrs. The choice of integration time of 120\,hrs is relatively conservative, as the EoR0 field will potentially be observed for $>1000$\,hrs. We note that, a large amount of data loss is expected for future surveys due to radio frequency interference (RFI) flagging. It is also common in data analysis to cross-correlate different time blocks to reduce systematic effects (e.g. \citealt{2022ApJ...925..221A,2023arXiv230111943P}), which effective reduces the integration time by half. Overall, we expect $t_{\rm tot} = 120\,$hrs to be a reasonable lower limit for future SKA-Low observations of one deep field.

As a preliminary study into the detectability and constraining power of EoR multipoles, simulating observational systematics is beyond the scope of our work. In particular, we assume that foreground contamination can be avoided by applying a horizon criterion so that
\begin{equation}
    k_\parallel > c_k k_\perp,
\label{eq:fgcriterion}
\end{equation}
where the position of the wedge $c_k$ is determined by a characteristic angular scale $\theta_0$ so that \citep{2014PhRvD..90b3018L}
\begin{equation}
    c_{k} = \frac{H(z)D_{c}(z)\theta_0}{c(1+z)},
\label{eq:horizon}
\end{equation}
where $D_c(z)$ is the comoving distance. We choose $\theta_0 = \sqrt{\Omega_{\rm beam}}$ following \cite{2023MNRAS.524.3724C} which gives $c_k = 0.28$, where $\Omega_{\rm beam}$ is the beam area calculated using the power beam shown in \autoref{fig:beam}. As the 21\,cm signal during the EoR is anisotropic, the power spectrum averaged above the foreground wedge is biased compared to the spherical average \citep{2016MNRAS.456...66J}. It also leads to bias in the multipoles, inducing effective mode-mixing in the 1D multipole power spectra \citep{2018MNRAS.475..438R}. In this paper, we refer to the average above the foreground wedge as ``spherical average'' for simplicity. The partition into clustering wedges discussed in \hyperref[subsec:nonuniform]{Section \ref{subsec:nonuniform}} is performed above the foreground wedge instead of using the entire $\bm{k}$-space.  {\autoref{eq:fgcriterion} can also be written as}
\begin{align}
    \mu > \mu_{\rm fg}, \\
    \mu_{\rm fg} = \frac{c_k}{\sqrt{1+c_k^2}},
\end{align}
 {where $\mu \equiv k_\parallel / |\bm{k}|$ is used for the multipole expansion discussed later in \hyperref[sec:psmulti]{Section \ref{sec:psmulti}}}.

Several observational effects induce foreground leakage into the observation window above the wedge. For example, chromatic data excision due to narrowband RFI will cause foreground spillover \citep{2022MNRAS.510.5023W}. Calibration errors also introduce foreground scatter into high delay (e.g. \citealt{2016MNRAS.461.3135B,2019ApJ...875...70B}). Although not discussed in this paper, we note that the systematic effects can be partially mitigated in the quadratic estimator formalism, for example by including the foreground cleaning operation in the weighting matrix $\mathbf{R}$ (e.g. \citealt{2021MNRAS.501.1463K,2023MNRAS.518.2971C,2023MNRAS.524.3724C}).

\section{Power spectrum multipoles in clustering wedges}
\label{sec:psmulti}
 {In this section, we give an introduction to the power spectrum multipoles in clustering wedges as summary statistics of the EoR. In particular, we discuss the relation between the multipoles and the anisotropy of the 3D power spectrum, and the observational effects in the measured multipole power spectra. These effects lay the foundations for the incentives for using multipoles in wedges, which we discuss later in \hyperref[sec:compare]{Section \ref{sec:compare}}}.

\subsection{Definitions}
In the flat-sky and plane-parallel limit, the power spectrum multipole of order $\ell$ in a clustering wedge can be written as (e.g. \citealt{2015PhRvD..92h3532S})
\begin{equation}
    P^\ell_{21} (k) = \frac{2\ell+1}{\mu_1-\mu_0}\int^{\mu_1}_{\mu_0} {\rm d}\mu \, \mathcal{P}_\ell(\mu)\, w(\bm{k}) \, P_{21}(\bm{k}),
\label{eq:multipole}
\end{equation}
where $k = |\bm{k}|$ is the mode of the 3D wavenumber vector, $\mu = k_\parallel/k$, $\mu_0$ and $\mu_1$ are the lower and upper limits of the clustering wedge, $\mathcal{P}_\ell$ is the Legendre polynomial, $P_{21}(\bm{k}) \equiv P_{21}(k,\mu)$ is the anisotropic 21\,cm power spectrum and $w(\bm{k})$ is the weighting of the 3D power spectrum.  {The weights are renormalised so that for a specific 1D $k$-bin, $\int {\rm d}\mu \, w(k,\mu) = 1$}. We choose the spacing of the $k$-bins to be logarithmically distributed from $0.05\,{\rm Mpc^{-1}}$ to $1\,{\rm Mpc^{-1}}$ with 20 bins. We consider $\ell = 0,2,4$, i.e. the monopole, quadrupole and hexadecapole, in this paper.

For illustration, we also use the ``dimensionless'' power spectrum \footnote{In the case of the 21\,cm power spectrum, $\Delta^\ell_{21} (k)$ has the unit of [T$^2$]. We use the term ``dimensionless power spectrum'' analogous to the dimensionless density power spectrum. }
\begin{equation}
    \Delta^\ell_{21} (k) = \frac{k^3}{2\pi^2} P^\ell_{21}(k).
\label{eq:psdimless}
\end{equation}

 {From \autoref{eq:multipole}, we can introduce the terminology of the summary statistics. A ``wedge'' is defined as a region in the $\bm{k}$-space enclosed by the integration limit $\mu_0 < \mu < \mu_1$. For example, the region dominated by the foregrounds, as defined in \autoref{eq:horizon}, is referred to as the foreground wedge and corresponds to $\mu_0 = 0$, $\mu_1 = \mu_{\rm fg}$. Consequently, a ``clustering wedge'' is a region in the $\bm{k}$-space outside the foreground wedge with an arbitrary choice of $\mu_0$ and $\mu_1$, as long as $\mu_0 > \mu_{\rm fg}$. Throughout this paper, we use ``spherical average'' to refer to the case where there is no further splitting of $\bm{k}$-space above the foreground wedge. On the contrary, we use ``monopole/multipoles in clustering wedges'' to refer to the case where the region above the foreground wedge is split into multiple wedges.}

\subsection{Multipoles and anisotropy}
 {Using the formalism described by \autoref{eq:multipole}, we first discuss the relation between the power spectrum anisotropy and the multipoles, without the partitioning of clustering wedges and observational effects. In this case, the integration range of $\mu$ in \autoref{eq:multipole} is simply $\mu_1 = 1$ and $\mu_0 = 0$. The power spectrum is uniformly weighted so that $w(\bm{k}) = 1$. We note that the foreground wedge is not included in this subsection and will be discussed later in \hyperref[subsec:obseffects]{Section \ref{subsec:obseffects}}.}

\begin{figure*}
    \centering
    \includegraphics[width=\linewidth]{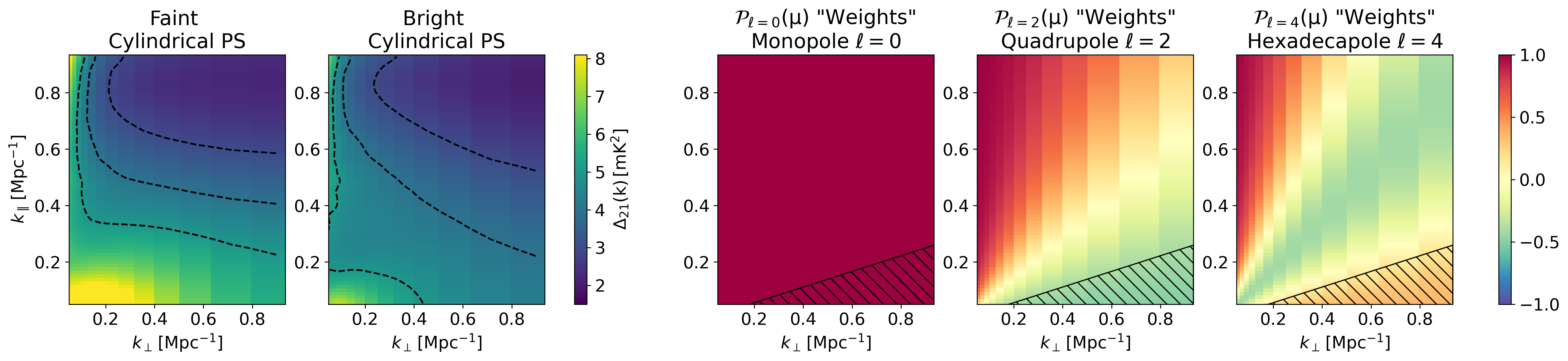}
    \caption{ {Left panels: The dimensionless cylindrical power spectrum of the 21\,cm signal of the faint and bright models. The dashed lines correspond to the contours where $\Delta_{21}(\bm{k}) = [3,4,5]\,[{\rm mK^2}]$. The results are averaged over 10 realizations. Right panels: The values of the Legendre polynomials $\mathcal{P}_{\ell}(\mu)$ for $\ell = 0,2,4$. The regions shaded by slashes correspond to the foreground wedge.}}
    \label{fig:multiillus}
\end{figure*}

 {From \autoref{eq:multipole}, it can be seen that multipoles of different order $\ell$ are essentially 1D averages of the 3D power spectrum with different weighting schemes, and the weighting schemes are described by the Legendre polynomial $\mathcal{P}_{\ell}(\mu)$. As a result, the weighted averages of the power spectrum can be used as a probe for the anisotropy, as illustrated in \autoref{fig:multiillus}. For both faint and bright models, it can be seen that the signal power spectrum is highly anisotropic, demonstrated by the contours of constant power spectrum amplitude in the left panels of \autoref{fig:multiillus}. The amplitude of the 21\,cm power spectrum is visibly higher around $k_\parallel \sim 1\,{\rm Mpc^{-1}}, k_\perp \sim 0$. This is due to the fact that the power spectrum is measured along a large lightcone shown in \autoref{fig:lightcone}. Along the line-of-sight direction, there is a large evolution of the global ionization fraction of the IGM, which corresponds to an enhanced fluctuation of at large transverse, small radial scales. Meanwhile, the power spectrum amplitude is also higher around $k_\parallel \sim 0, k_\perp \lesssim 0.5\,{\rm Mpc^{-1}}$. At these scales, the power spectrum probes the distribution of ionization sources averaged along the line-of-sight. Overall, for the same $|\bm{k}|$, the power spectrum has higher amplitude at smaller $\mu$.}

 {The morphology of ionization bubbles and their evolution lead to the anisotropy in the power spectrum. As a result, the anisotropic features in the power spectrum differ for the faint and bright models, as seen in \autoref{fig:multiillus}. For the bright model, the amplitude of the power spectrum is significantly lower at $k_\parallel \sim 0, k_\perp \lesssim 0.5\,{\rm Mpc^{-1}}$ comparing to the faint model. As shown in \autoref{fig:lightcone}, the bright model gives a more rapid reionization process, with larger but fewer bubbles. Therefore, the fluctuation of the brightness temperature is smaller at small transverse scales for the bright model.}

\begin{figure}
    \centering
    \includegraphics[width=\linewidth]{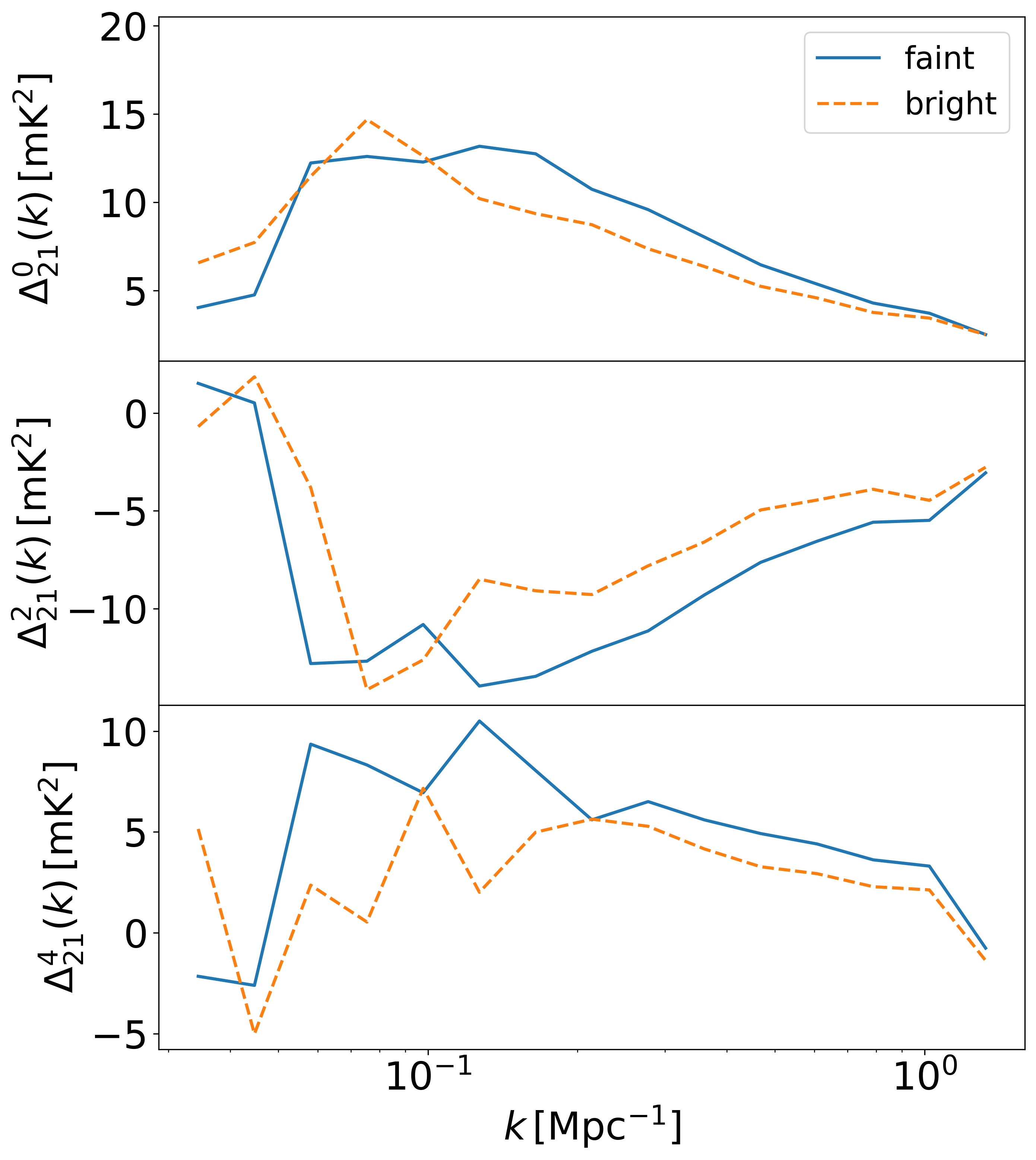}
    \caption{
     {The dimensionless power spectrum multipoles $\Delta^{0,2,4}_{21} (k)$ {averaged over 10 realizations} for both the faint and bright model. The multipoles are integrated from $\mu_0 =0 $ to $\mu_1 = 1$ with uniform weighting $w(\bm{k})=1$ as defined in \autoref{eq:multipole}.} 
    }
    \label{fig:multisimple1d}
\end{figure}

 {The differences in the anisotropy are then captured in the power spectrum multipoles. In \autoref{fig:multisimple1d}, we show the averaged 1D multipole power spectrum of the 21\,cm signal. The monopole power spectrum gives constant weights $\mathcal{P}_{\ell=0}(\mu)=1$ to the 3D power spectrum before performing the 1D averaging. Therefore, the monopole power spectrum only gives the overall amplitude averaged along a certain scale $|\bm{k}|$. On the other hand, the quadrupole gives positive weights to large $\mu$ and negative weights to small $\mu$. As the 21\,cm power spectrum is lower at higher $\mu$, the overall quadrupole is negative except for large scales $k\lesssim 0.05\,{\rm Mpc^{-1}}$. The hexadecapole weight $\mathcal{P}_{\ell=4}(\mu)$ is positive at $\mu \sim 0$. It first decreases as $\mu$ increases and the increases, leading to negative values at $\mu \sim 0.5$ and positive values at $\mu \sim 1$. The averaging results in a positive amplitude in the 1D average. Since the power spectrum is less anisotropic for the bright model, the amplitudes of the multipoles are lower comparing to the faint model. This suggests that multipoles can be used to extract information on the reionization model, as we discuss more in \hyperref[subsec:info]{Section \ref{subsec:info}}.}

\subsection{Observational effects in multipoles}
\label{subsec:obseffects}
 {The 21\,cm multipoles reflect the underlying anisotropy of the 3D power spectrum. However, the multipoles are also affected by observational effects. The observational effects influence the summary statistics in different ways, which we outline in this subsection.}

\begin{itemize}
    \item 
    There is mode-mixing along the radial and the transverse direction. Along the line-of-sight, different $k_\parallel$-modes are mixed due to bandpass tapering. $\bm{k}_\perp$-modes are also mixed due to the attenuation of the primary beam. In this work, we assume mode-mixing is negligible by using relatively large $u$-$v$ grids and a renormalisation matrix in the quadratic estimator for power spectrum estimation. See \hyperref[apdx:est]{Appendix \ref{apdx:est}} for more discussions.
    \item
    The foreground wedge forbids measurements in $\mu<\mu_{\rm fg}$. In the ideal case, the entire $\bm{k}$-space from $\mu=0$ to $\mu=1$ is sampled. For spherically averaged multipoles, the measured spectra integrated from $\mu_0 = \mu_{\rm fg}$ will be biased comparing to the ideal spectra integrated from $\mu_0 = 0$.
    \item 
    The $\bm{k}$-space is not uniformly sampled, with an additional weighting introduced by the baseline distribution. In the ideal case, the weighting of the $\bm{k}$-modes $w(\bm{k})$ is uniform so that $w(\bm{k}) = 1$. In real observations, however, the $\bm{k}$-modes will be weighted by the inverse of the noise covariance, so that $w(\bm{k}) = \frac{{\rm d}N}{{\rm d}\bm{k}_\perp}$ where $\frac{{\rm d}N}{{\rm d}\bm{k}_\perp}$ is the probability distribution of the baselines.
\end{itemize}

\begin{figure}
    \centering
    \includegraphics[width=\linewidth]{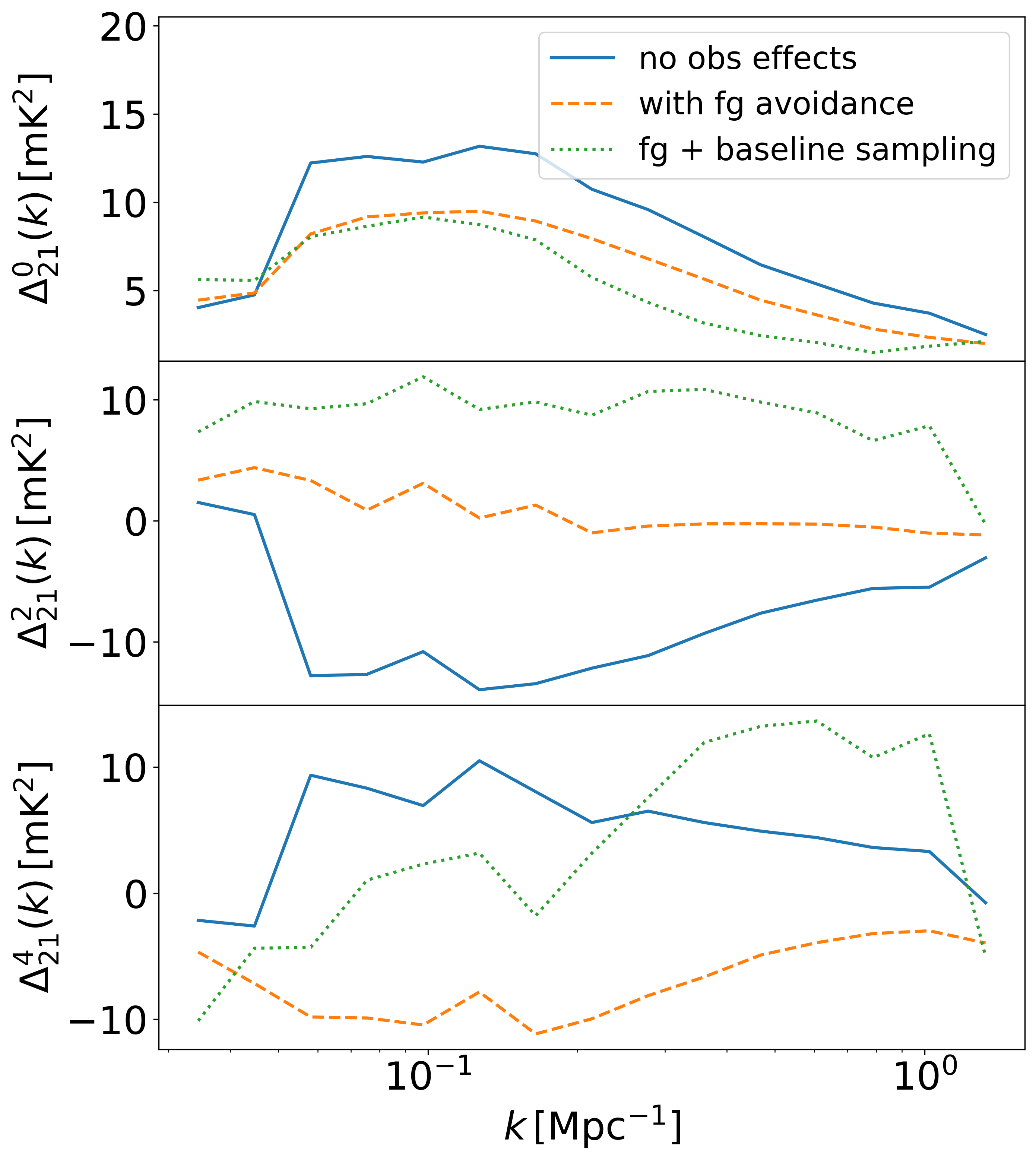}
    \caption{
     {The dimensionless power spectrum multipoles $\Delta^{0,2,4}_{21} (k)$ {averaged over 10 realizations} for the fiducial faint model, with no foreground contamination and uniform weighting (``no obs effects''), foreground avoidance and uniform weighting (``with fg avoidance''), and foreground avoidance and weighting based on baseline distribution (``fg + baseline sampling'').} 
    }
    \label{fig:psbias}
\end{figure}

 {These observational effects can be disentangled from the underlying 21\,cm signal by utilising the multipoles in clustering wedges, which we discuss in detail in \hyperref[sec:compare]{Section \ref{sec:compare}}. For this section, we first establish the intuition of the importance of these effects, by showing the biasing of the power spectrum from the observational effects in \autoref{fig:psbias}. For simplicity, only the faint model is shown here. When foreground avoidance is included, the monopole power spectrum decreases in amplitude. This is due to the fact that the lower $\mu$ region is excluded where the power spectrum amplitude is higher. For the quadrupole, the foreground wedge corresponds to the region where the Legendre polynomial $\mathcal{P}_{\ell = 2}(\mu)$ is negative. Therefore, excluding the foreground wedge results in a positive-valued quadrupole power spectrum. For the hexadecapole, when foreground avoidance is not included, the averaging gives positive amplitude since it weights $\mu\sim 0$ positively. Including the foreground avoidance gives negative amplitude, since $\mathcal{P}_{\ell = 4}(\mu)$ is negative for $\mu\sim 0.5$ where the 3D power spectrum has higher amplitude, while positive for $\mu \sim 1.0$ where the 3D power spectrum is lower.}

 {The uneven $\bm{k}_\perp$ sampling further biases the power spectrum multipoles. The exact effects on the multipoles depend on the baseline distribution and $|\bm{k}|$-bins. For SKA-Low, we find that in general, higher values of $\mu$ are sampled more frequently, which we discuss further in \hyperref[subsec:nonuniform]{Section \ref{subsec:nonuniform}}. The monopole power spectrum is therefore lowered as its amplitude is smaller at large $\mu$. On the other hand, the higher-order Legendre polynomials are positive at $\mu\sim1$. Therefore, the quadrupole and the hexadecapole become larger when baseline distribution is considered. }

 {In conclusion, the observational effects have significant impact on the measured power spectrum multipoles of the 21\,cm signal. In turn, measuring the $(\bm{k}_\perp,k_\parallel)$ dependency of the power spectrum amplitude helps disentangle observational effects with the underlying signal. This provides strong incentives to use power spectrum multipoles in clustering wedges, which we demonstrate in the next section.}

\section{Advantages of multipoles and clustering wedges}
\label{sec:compare}
In this section, we demonstrate the advantages of using multipoles in clustering wedges as summary statistics for EoR surveys. Using power spectrum multipoles allows us to extract information from anisotropy and to probe the evolution of the EoR along the lightcone. Furthermore, the advantage of partitioning the measurements into clustering wedges is illustrated. For simplicity, when not specifically mentioned, the results shown correspond to the simulation using the faint model defined in \hyperref[subsec:21cmsim]{Section \ref{subsec:21cmsim}}.  {The results shown are averages over 10 independent realizations. From here on, all power spectra are calculated with the inverse noise covariance weighted, baseline-sampled 3D bandpowers using the quadratic estimator formalism described in Appendix \ref{apdx:est}. }

In order to understand the information content of the multipoles, we also need to compute the covariance of the power spectra. While the noise covariance can be analytically calculated  {in terms of its two-point correlation function (see Appendix \ref{apdx:est}), the signal itself is non-Gaussian and requires the computation of high-order correlation functions \citep{2020MNRAS.498.1480S} to calculate its covariance. Instead of the quadratic estimator,} we use the jackknife method and calculate the signal covariance directly from the simulation lightcone. The lightcone is resampled 25 times along the transverse direction, each with a sub-area of $(320\,{\rm Mpc})^2$ taken out. The conversion from the delay power spectrum to the 21\,cm temperature power spectrum is susceptible to the lightcone effects, the plane-parallel approximation, and the treatment of the primary beam. For the results shown in this paper, we use the measured power spectra from visibility data, and ignore the systematic biases of the estimation. We note that, simulation-based inference (e.g. \citealt{2022ApJ...926..151Z,2023MNRAS.525.6097S,2024arXiv240314060G}) can be used to circumvent these problems, as it does not require an explicit likelihood and takes into account systematic effects through forward-modelling. The detailed treatment of an inference framework to be used on real data is beyond the scope of our work.

\subsection{Information on anisotropy}
\label{subsec:info}
\begin{figure}
    \centering
    \includegraphics[width=\linewidth]{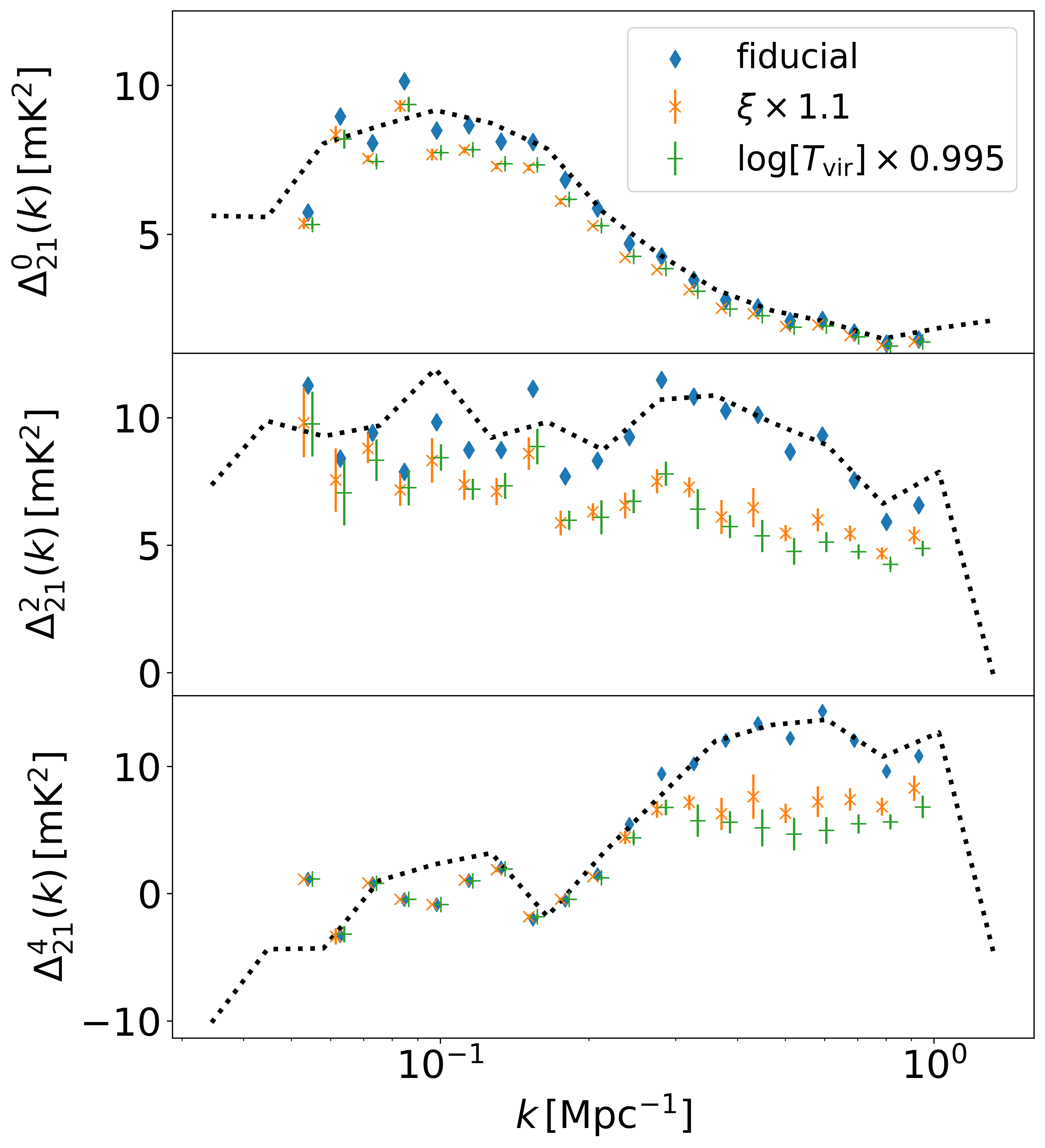}
    \includegraphics[width=\linewidth]{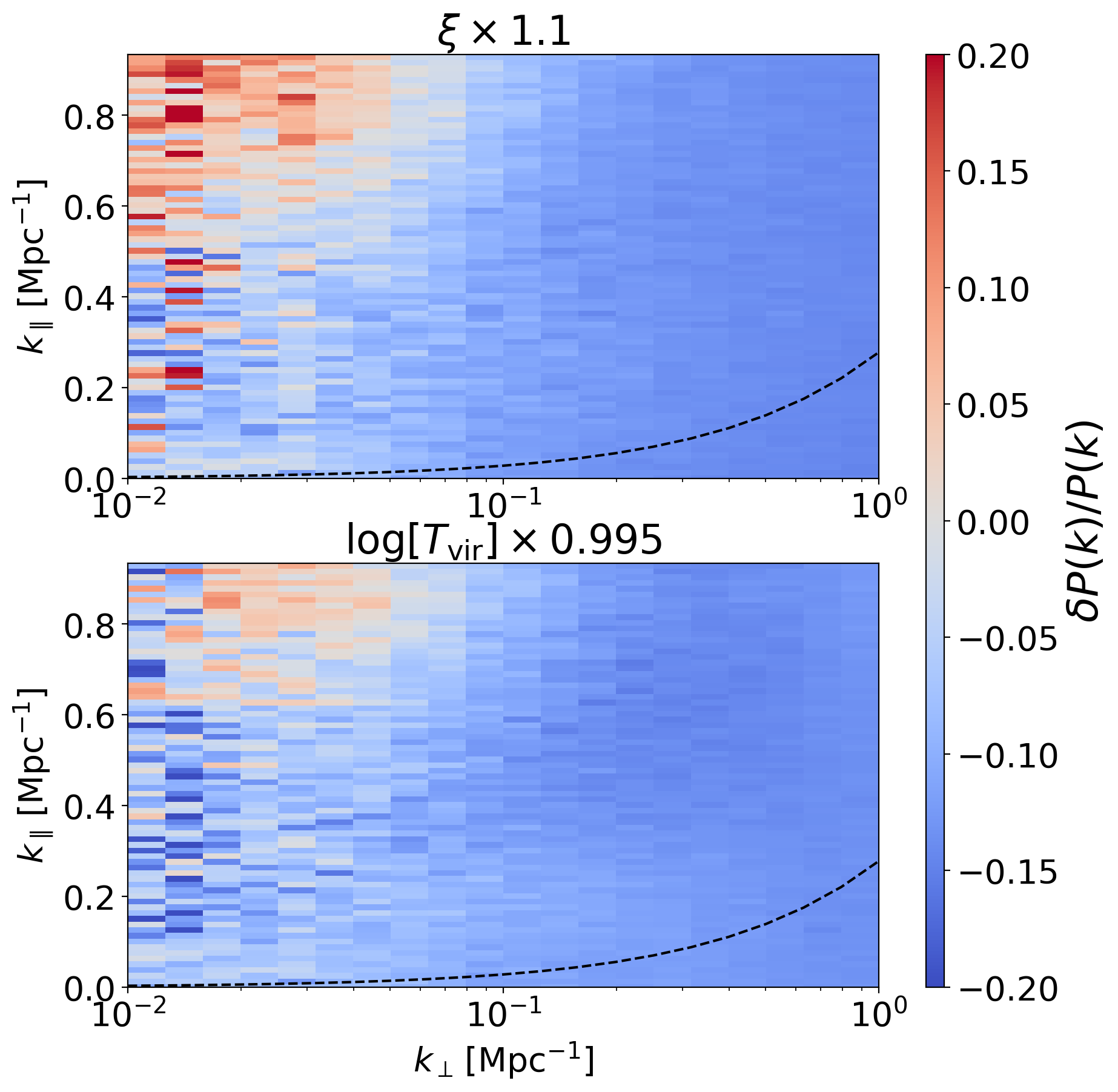}
    \caption{
    Top panel: The dimensionless power spectrum multipoles $\Delta^{0,2,4}_{21} (k)$  {averaged over 10 realizations} for the fiducial faint model ('fiducial'), the model with $\xi$ being 10\% higher than the fiducial value (``$\xi \times 1.1$''), and the model with $\log [T_{\rm vir}]$ being 0.5\% lower than the fiducial value (``$\log [T_{\rm vir}] \times 0.995$'').  {The error bars correspond to the standard deviation of the signal among the realizations.} The latter two cases show very similar amplitudes for $\Delta^{0}_{21} (k)$, suggesting a degeneracy between the model parameters in the 1D monopole ($\ell=0$). The degeneracy is broken at higher $\ell$. The power spectrum of the input signal is shown in black dotted line for reference.  {The $k$ values for ``$\xi \times 1.1$'' and ``$\log [T_{\rm vir}] \times 0.995$'' have been shifted by 2\% for better visualisation.}
    Bottom panel: The fractional differences in the cylindrical power spectrum for the ``$\xi \times 1.1$'' and ``$\log [T_{\rm vir}] \times 0.995$'' models compared to the fiducial. Values larger than 0.2 are set to 0.2 and values smaller than -0.2 are set to -0.2 for better visualization.
    }
    \label{fig:div_illus}
\end{figure}

The fluctuation of the 21\,cm signal, as described by \autoref{eq:tb}, is determined by the multiplication of the ionization field, the spin temperature, the matter density field and the velocity field. While quantities such as the velocity field (e.g. \citealt{2019MNRAS.490.1255C}) and the spin temperature (e.g. \citealt{2024arXiv240408042S}) have a sizeable impact on the 21\,cm power spectrum, the features of the fluctuation are mostly influenced by the matter density on large cosmological scales and the ionization field on small scales \citep{2022MNRAS.513.5109G}. Particularly, after the beginning of the percolation of the ionization bubbles, the ionization field becomes highly non-Gaussian (e.g. \citealt{2015MNRAS.451..467S}) and non-linear (e.g. \citealt{2016MNRAS.457.1550H}). As a result, the evolution of ionization bubbles along the line-of-sight induces anisotropy into the 21\,cm power spectrum. The anisotropy can be captured in the cylindrical power spectrum. However, observational effects such as beam attenuation and frequency-dependent sampling create mode-mixing of different $\bm{k}$-modes, making it difficult to accurately measure the signal and its covariance in fine cylindrical $\bm{k}$-bins. Alternatively, power spectrum multipoles can be used for measuring the anisotropy.

\begin{figure*}
    \centering
    \includegraphics[width=\linewidth]{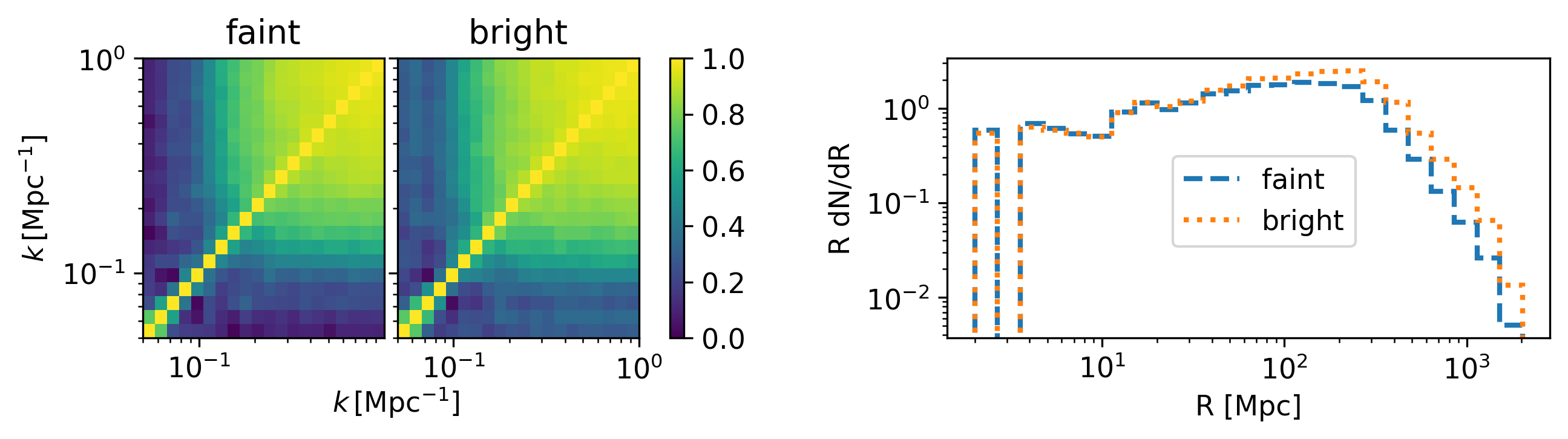}
    \caption{
    Left panel: The correlation matrix of the power spectrum monopole  {of the pure 21\,cm signal} for the faint and bright models.  {The non-uniform sampling of $k$-space due to the baseline distribution is propagated into the covariance calculation (see \hyperref[subsec:nonuniform]{Section \ref{subsec:nonuniform}}).}
    Right panel: Bubble size distribution of the ionization field lightcone the for faint and bright models. The bubble sizes are calculated using the mean-free-path method, and only completely ionized cells are considered.
    }
    \label{fig:corr_illus}
\end{figure*}

For the spherically averaged power spectrum monopole, anisotropic information is lost, leading to obscurity in the interpretation of the amplitude of the signal. The lack of distinguishing power in the 1D monopole is illustrated in the upper panel of \autoref{fig:div_illus}. For the cases of increasing $\xi$ and decreasing $\log T_{\rm vir}$, the changes in the power spectrum amplitude are similar. This leads to a potential degeneracy between the model parameters, and indicates the limitations in the constraining power from monopole. On the other hand, as shown in the lower panel of \autoref{fig:div_illus}, the differences in the cylindrical power spectrum contain information on the model parameters. The constraining power is then reflected in the higher order multipoles $\ell = 2$ and $\ell = 4$. Increasing $\xi$ and decreasing $\log T_{\rm vir}$ lead to changes in the multipole power spectra with different structures. The differences are visible at all scales in the quadrupole, which can also be seen in the hexadecapole at small scales $k\gtrsim 0.3\,{\rm Mpc^{-1}}$.

\subsection{Decorrelation of small-scale measurements}
\label{subsec:decorr}
The patchiness of ionization bubbles during the EoR breaks the tracing of the brightness temperature field to the density field. As a result, at small scales the temperature power spectrum is highly correlated towards the end of reionization (e.g. \citealt{2017MNRAS.464.2992M}). This is due to the fact that the signal is dominated by correlation within the same ionized structure at scales smaller than the typical size of the bubbles. 

To illustrate this correlation, we show the correlation matrix of the monopole  {of the pure 21\,cm signals} in \autoref{fig:corr_illus} for both the faint and bright models. For the faint model, the power spectrum signal is highly correlated at relatively small scales $k\gtrsim 0.2\,{\rm Mpc^{-1}}$. The monopole power spectrum is even more correlated for the bright model, especially on the large scales. The small-scale correlation indicates that the fluctuations of the ionization field are dominating the signal. To further establish the connection between the structures of the bubbles and the power spectrum correlation, we show the bubble size distribution for both faint and bright models in the right panel of \autoref{fig:corr_illus}. The bubble size distribution is calculated using the mean-free-path method \citep{2007ApJ...669..663M}, and only cells that are completely ionized are counted into the bubble sizes. Due to the fact that the reionization process is driven by brighter and less numerous sources in the bright model, there are more large bubbles in the bright model than in the faint model, which can be seen in \autoref{fig:corr_illus}. The increase in bubble sizes corresponds to the stronger correlation at large scales, verifying our conclusion on the connection between bubble sizes and signal correlation.

As the signal becomes completely correlated at small scales $k\sim 0.3\,{\rm Mpc^{-1}}$, the constraining power of the monopole on EoR parameters is severely limited. Although SKA-Low will be able to probe a wide range of scales, the extraction of information from measurements of the monopole is not sufficient. For forthcoming surveys, this issue may be compounded by the fact that shorter baselines are more susceptible to observational systematics (e.g. \citealt{2023ApJ...947...16A,2023arXiv230111943P}), leading to further signal loss. Alternative summary statistics that can probe small scales are thus called for.

\begin{figure}
    \centering
    \includegraphics[width=\linewidth]{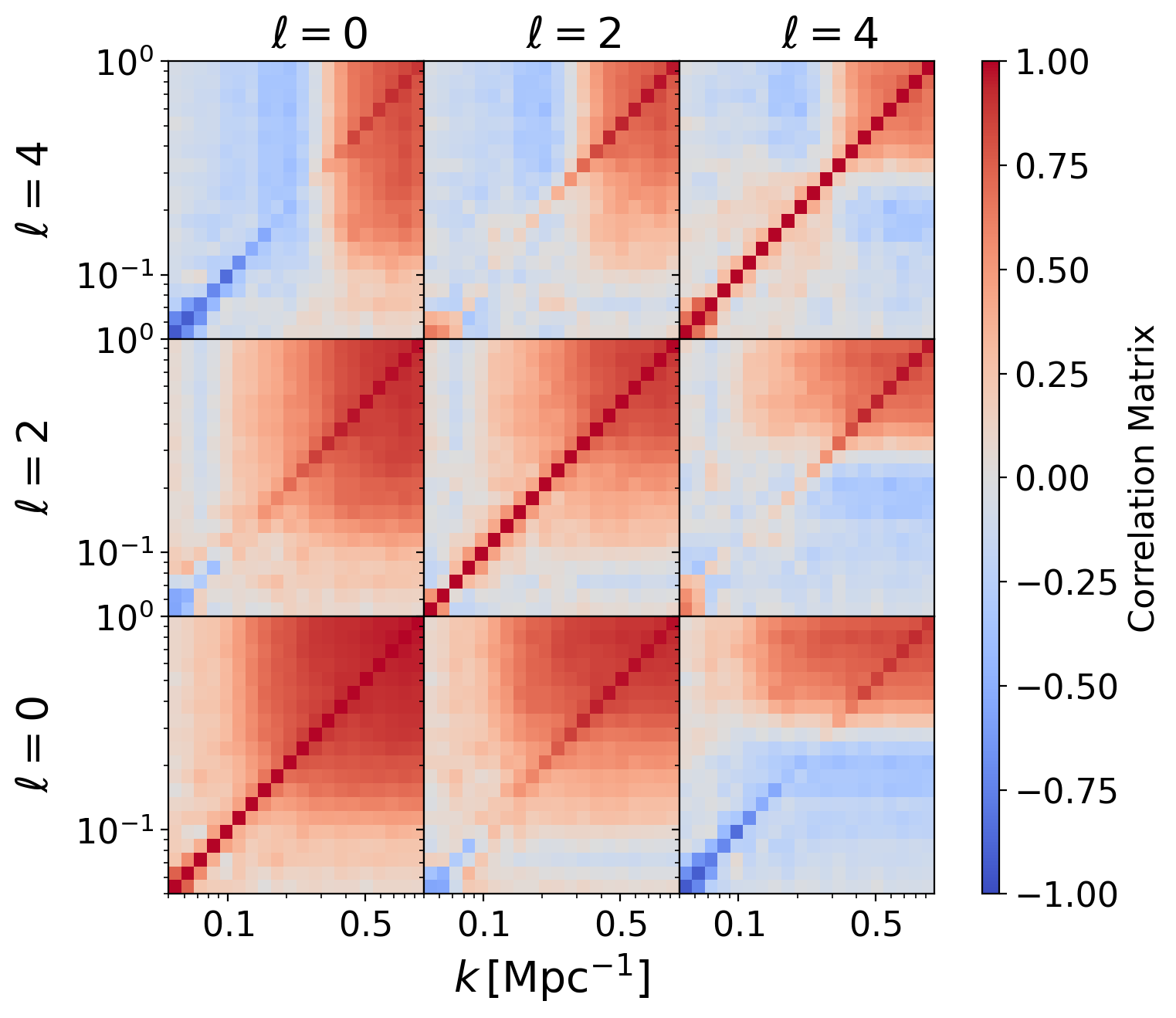}
    \caption{
    The correlation matrix of the power spectrum multipoles of the faint model. The signal is strongly correlated at $\ell = 0$ (the monopole) on small scales $k\gtrsim 0.2 \, {\rm Mpc^{-1}}$. The correlation becomes weaker at higher $\ell$.  {The non-uniform sampling of $k$-space due to the baseline distribution is propagated into the covariance calculation}.
    }
    \label{fig:corr_avg}
\end{figure}

In order to recover information from the small scale measurements, we can utilise higher-order multipoles. As we have shown in \hyperref[sec:psmulti]{Section \ref{sec:psmulti}}, the multipoles probe the evolution of the signal along the line-of-sight, subsequently disentangling the signal at smaller scales. In \autoref{fig:corr_avg}, we show the correlation matrix for the multipole data vector  {of the pure 21\,cm signal} including $\ell = 0,2,4$. For the quadrupole, the signal completely decorrelates at large scales $k\lesssim 0.1\,{\rm Mpc^{-1}}$ and becomes less correlated compared to the monopole at $k\gtrsim 0.1\,{\rm Mpc^{-1}}$. For the hexadecapole, the correlation is even weaker and the signal only becomes correlated at $k\gtrsim 0.5\,{\rm Mpc^{-1}} $. Combined with the fact that the multipoles can distinguish changes in the reionization parameters as shown in \autoref{fig:div_illus}, this indicates that including the quadrupole and the hexadecapole will significantly increase the information content of 21\,cm surveys.

The cross-correlation between $P^{\ell = 0}_{21}$ and $P^{\ell = 4}_{21}$ exhibits a transition from anticorrelation to correlation, as shown in \autoref{fig:corr_avg}. This is due to the baseline distribution leading to different sampling of the cylindrical $k$-space at different scales, which we discuss next. 

\subsection{Non-uniform sampling from baseline distribution}
\label{subsec:nonuniform}
The $\bm{k}$-modes measured in a 21\,cm survey are determined by the distribution of the interferometric baselines. For redundant arrays like HERA, specific $|\bm{k}_\perp|$ modes will be densely sampled and the $k$-dependence of the measured power spectrum is mainly contributed by the line-of-sight direction (see e.g. \citealt{2017PASP..129d5001D}). SKA-Low will be able to cover various scales on the transverse plane. However, the sampling of the $\bm{k}_\perp$ space is not homogeneous, as the sensitivity of the array will concentrate around specific scales of interest. As a result, the 1D average of multipoles will have highly uneven sampling across the cylindrical $\bm{k}$-space.

\begin{figure}
    \centering
    \includegraphics[width=\linewidth]{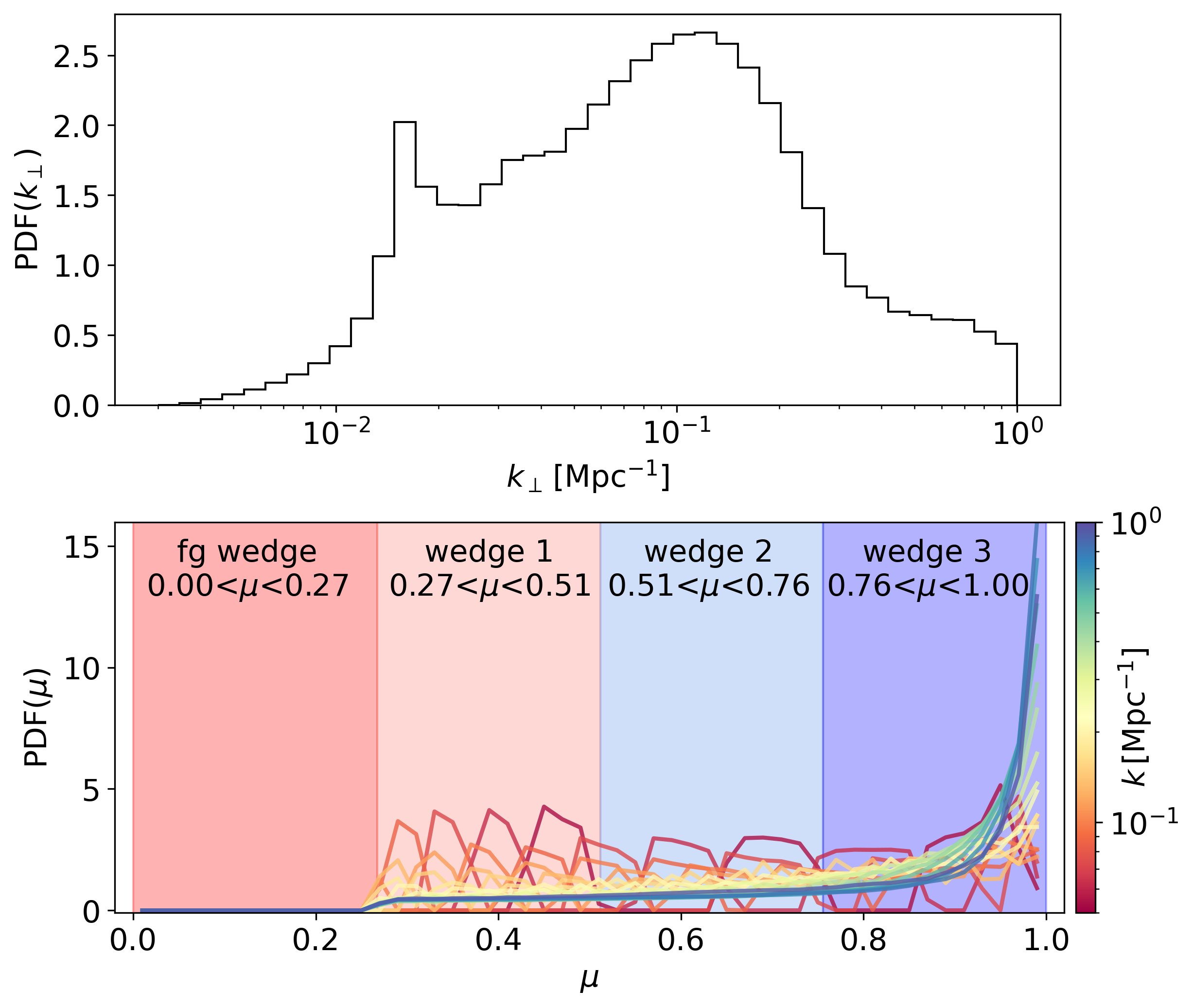}
    \caption{
    Top panel: The distribution function of the simulated SKA-Low baselines. The density of the baseline distribution peaks around $k_\perp \sim 0.1\,{\rm Mpc^{-1}}$.
    Bottom panel: The distribution function of $\mu$ in each 1D $k$-bin for spherically averaged power spectrum measurements. The distributions are colour-coded according to the $k$ value of each $k$-bin. The partitions of $\mu$-space into three different wedges are overlaid for reference. The foreground wedge is also shown. For large $k$, the sampling is skewed towards higher values of $\mu$ and is dominated almost entirely by $\mu \sim 1$ for $k\gtrsim 0.5\,{\rm Mpc^{-1}}$.
    }
    \label{fig:mu_sample}
\end{figure}

\begin{figure*}
    \centering
    \includegraphics[width=0.32\linewidth]{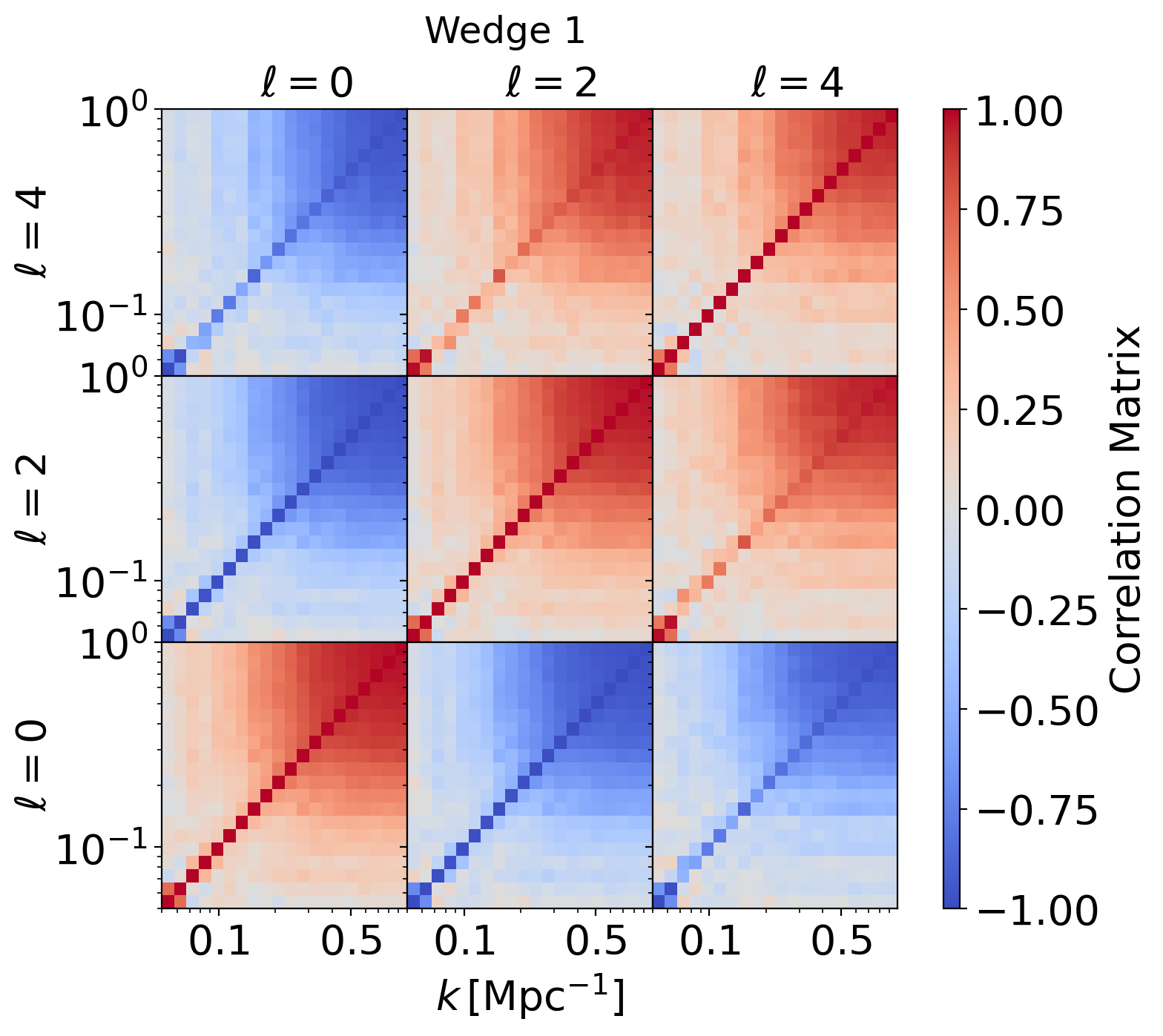}
    \includegraphics[width=0.32\linewidth]{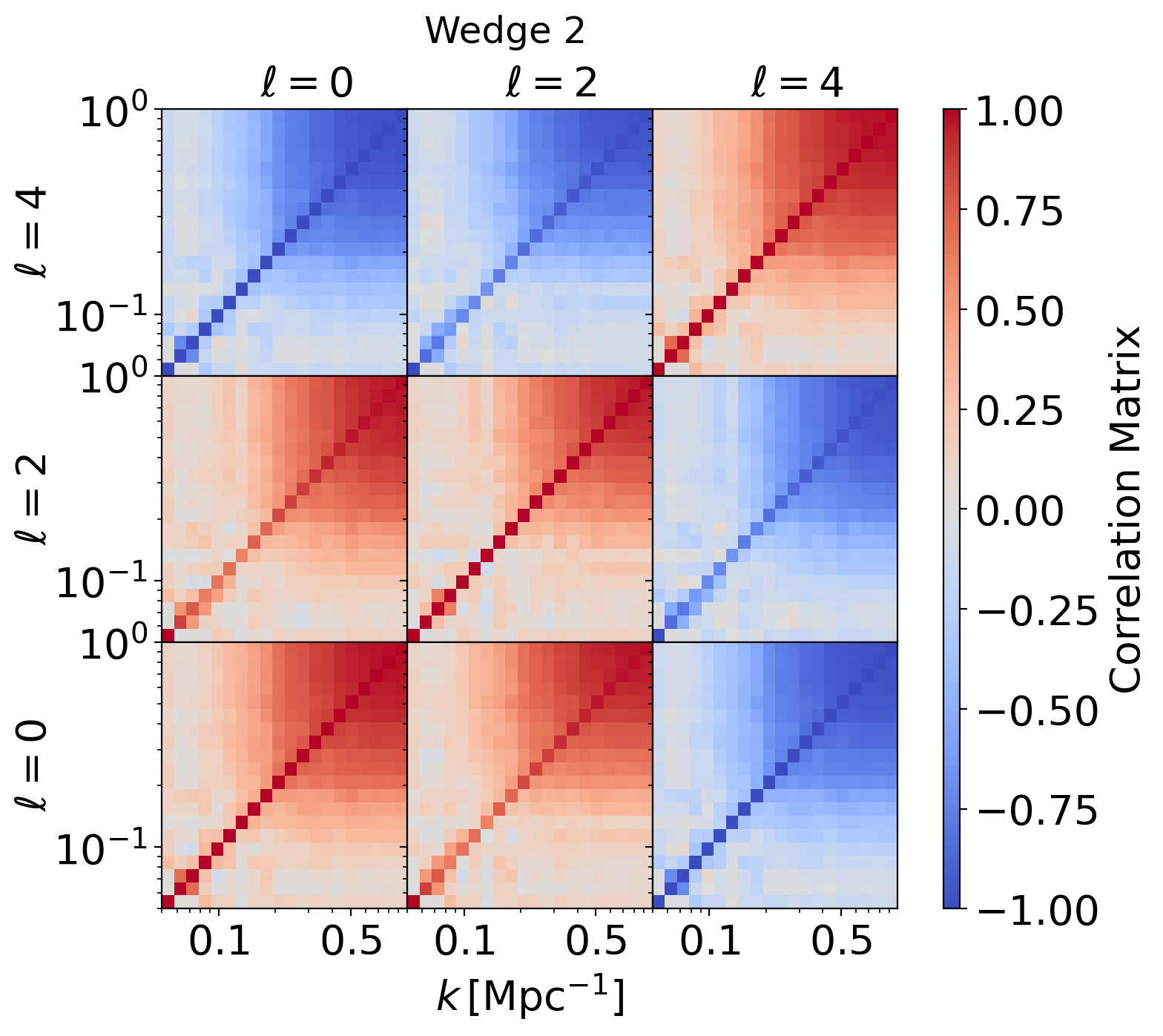}
    \includegraphics[width=0.32\linewidth]{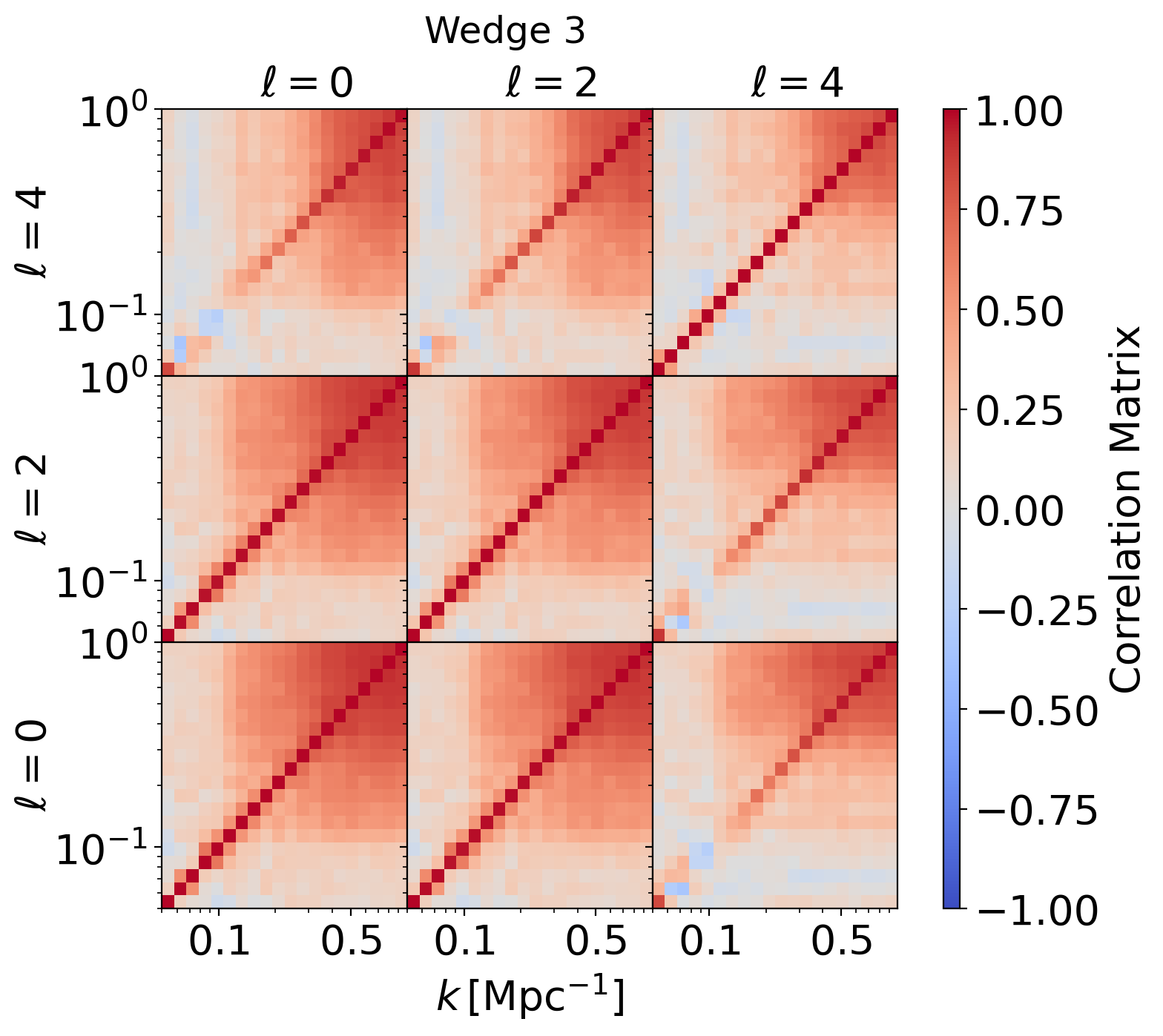}
    \caption{
    The correlation matrix of power spectrum multipoles  {of the pure 21\,cm signal} for different clustering wedges. The numbering and positioning of the wedges are defined in \hyperref[subsec:nonuniform]{Section \ref{subsec:nonuniform}} and shown in \autoref{fig:mu_sample}.  {The non-uniform sampling of $k$-space due to the baseline distribution is propagated into the covariance calculation}.
    }
    \label{fig:corr_wedge}
\end{figure*}

In \autoref{fig:mu_sample}, we show the 1D baseline distribution and the corresponding distribution function of $\mu$ for each 1D $k$-bin. The baseline distribution peaks around $k_\perp \sim 0.1\,{\rm Mpc^{-1}}$. Therefore, the sampling of $\mu$ is relatively even for large scales $k \lesssim 0.1\,{\rm Mpc^{-1}}$. On the other hand, the baseline distribution decreases sharply after the peak. For relatively small scales $k \gtrsim 0.1\,{\rm Mpc^{-1}}$, the contribution into the 1D power spectrum comes mainly from the large $k_\parallel$ region, skewing the distribution of $\mu$ towards higher values. The large deviation from uniformly sampled $\mu$ indicates loss of information, as spherically averaged power spectra will not be able to fully probe the $k_\parallel$-dependence of the signal and rather measures certain regions of large $\mu$.

To resolve the uneven sampling, we divide the $\mu$-region above the foreground wedge into three different clustering wedges, shown in \autoref{fig:mu_sample}. The foreground wedge is at $c_k = 0.28$ and equivalently $\mu = 0.267$. The $\bm{k}$-space above the foreground wedge is then split into three equally spaced regions of $\mu$. The partitioning resolves the uneven sampling by isolating less-sampled regions of $\mu$, leading to relatively even sampling in each clustering wedge. As mentioned in \hyperref[subsec:decorr]{Section \ref{subsec:decorr}}, one of the consequences of the uneven sampling is the transition from correlation to anticorrelation between the monopole and the hexadecapole as shown in \autoref{fig:corr_avg}. As a comparison, we show the correlation matrix for the multipoles  {of the pure 21\,cm signal} within each wedge in \autoref{fig:corr_wedge}. The transition disappears when the clustering wedges are used. We note that, due to the non-Gaussianity of the 21\,cm signal leading to higher-order correlation (e.g. \citealt{2018MNRAS.476.4007M}), on small scales the $k$-modes are correlated between clustering wedges. The full correlation matrix including the cross-correlation between different wedges is shown in \autoref{fig:corr_full} for reference. 

Comparing the correlation of multipoles in wedges with the spherical average, we can see that in the wedge with the largest $\mu$, the multipole power spectra are less correlated. This is consistent with the fact that the $\bm{k}$-modes with higher $\mu$ probe small line-of-sight scales, providing more information on the evolution along the lightcone. In the spherical average, at lower $k$ the sample is uniform. At higher $k$, the sampling becomes heavily skewed towards the third wedge, and the difference in the sampling of $\mu$ at different $k$ leads to the structure in \autoref{fig:corr_avg}.

\begin{figure}
    \centering
    \includegraphics[width=\linewidth]{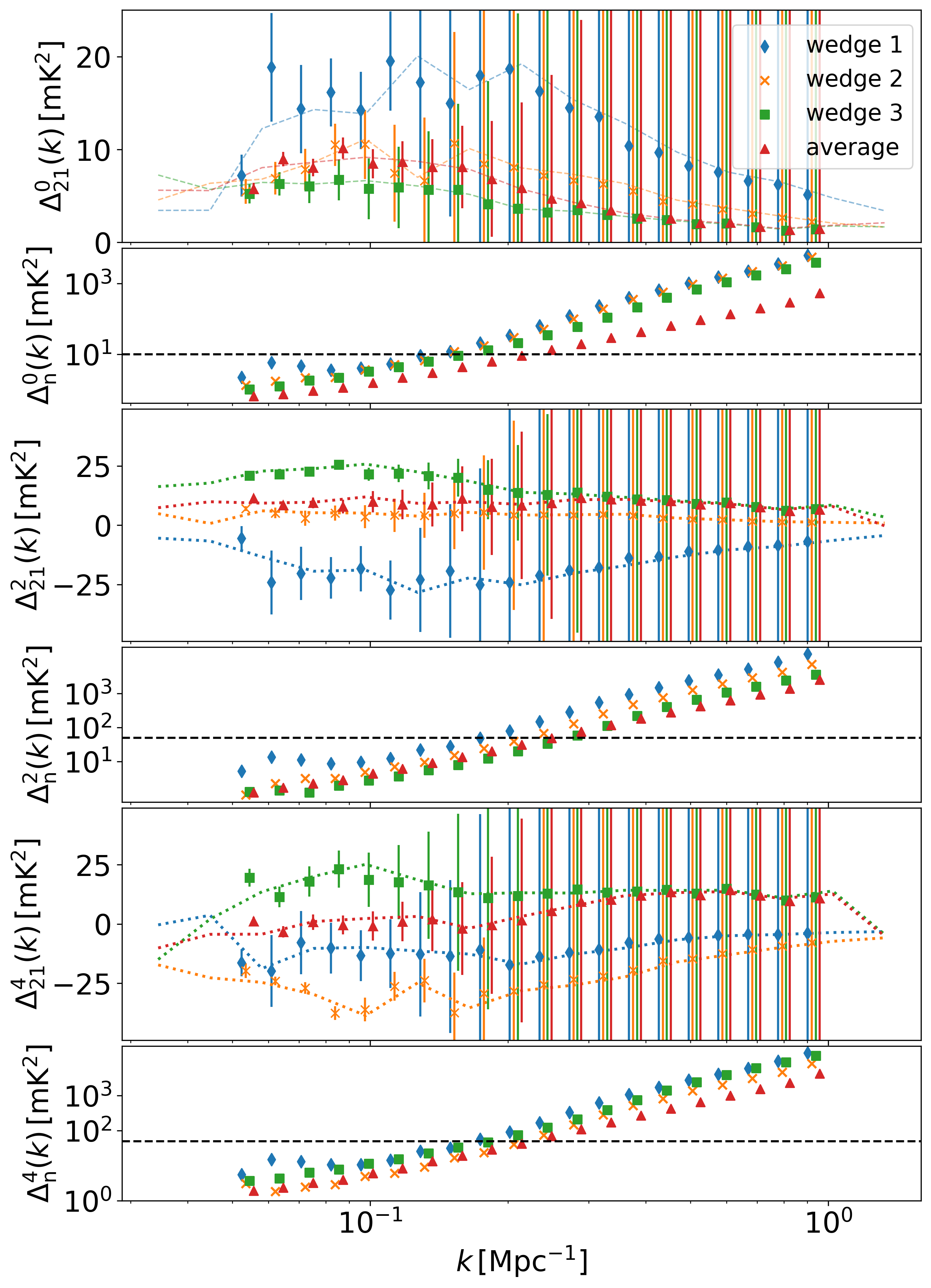}
    \caption{
    The forecasted measurements of the power spectrum multipoles for SKA-Low observations with 120\,hrs of integration time. The panels from top to bottom show the measurements for $\ell = 0$, $\ell = 2$ and $\ell = 4$. Measurements of each clustering wedge defined in \hyperref[subsec:nonuniform]{Section \ref{subsec:nonuniform}} are shown together with the measurements of the spherically averaged power spectrum (``average''). The dotted lines denote the signal power spectrum with no errors. In the lower part of each panel, the amplitudes of the measurements errors, denoted as $\Delta^{\ell}_{n}$, are also shown for reference. The black dashed line denotes a schematic upper limit for the noise level, above which the measurements no longer contain information. The centres of the $k$-bins are slightly shifted for better visualization.
    }
    \label{fig:forecast}
\end{figure}

\section{Parameter Constraints}
\label{sec:pars}
\subsection{Fisher matrix}
\label{subsec:fisher}
In this section, we forecast the constraining power of power spectrum multipoles for SKA-Low. The information content in summary statistics can be quantified via the Fisher matrix formalism (see e.g. \citealt{2020A&A...642A.191E})
\begin{equation}
    \mathbf{F}_{\alpha\beta} = \frac{1}{2}{\rm tr}\bigg[ 
    \frac{\partial \mathbf{C}}{\partial \theta_\alpha}\mathbf{C}^{-1} \frac{\partial \mathbf{C}}{\partial \theta_\beta}\mathbf{C}^{-1} 
    \bigg] + \frac{\partial \bm{\mu}^{\rm T}}{\partial \theta_\alpha} \mathbf{C}^{-1} \frac{\partial \bm{\mu}}{\partial \theta_\beta},
\label{eq:fisher}
\end{equation}
where $\rm tr$ denotes the trace of a matrix, $\bm{\mu}$ is the ensemble average of the data vector of the summary statistics, $\mathbf{C}$ is the data covariance, $^{\rm T}$ denotes the transpose operation and $\{\theta\}$ is the parameter set. The inverse of the Fisher matrix gives a lower bound of the error covariance of parameter inference \citep{a61aa5fe-d74a-3133-bed2-f35c3c555015,Rao1992}
\begin{equation}
    \mathbf{C}_{\theta} \geq \mathbf{F}^{-1},
\end{equation}
 {where $\geq$ indicates that $\mathbf{C}_{\theta} - \mathbf{F}^{-1}$ is positive semi-definite.}

The $1\sigma$ measurement error of a parameter $\alpha$ can then be written as
\begin{equation}
    \sigma_{\alpha} \geq \sqrt{(\mathbf{C}_{\theta})_{\alpha\alpha}}.
\end{equation}
The correlation coefficient between two parameters $\alpha$ and $\beta$ can be written as
\begin{equation}
    \rho_{\alpha\beta} = \frac{(\mathbf{C}_{\theta})_{\alpha\beta}}{\sqrt{(\mathbf{C}_{\theta})_{\alpha\alpha}(\mathbf{C}_{\theta})_{\beta\beta}}}.
\end{equation}
 {In the limit where the lower bound is achieved, we can use the inverse of the Fisher matrix as the covariance matrix of the parameters which we use for the forecast.}

To calculate the partial derivative of the power spectrum multipoles and the data covariance, we run the EoR simulations with small changes to the input parameters, shifting the values of the parameters around the fiducial values. We fix the initial condition when varying the EoR parameters in each realization, and average the calculated derivatives over 10 different realizations, after which we find that convergence has been reached  {as shown in \hyperref[apdx:convergence]{Appendix \ref{apdx:convergence}}.} We test the choice of step size between $0.1\%$ to $5\%$ and choose it to be $1\%$ to calculate the derivatives. While we find the correlation between the parameters vary slightly, the forecasts for parameter constraints discussed in \hyperref[sec:pars]{Section \ref{sec:pars}} are consistent with different step sizes.

In large scale structure surveys, it is often assumed that either the mean of the data vector is zero or the derivatives of the data covariance is negligible (e.g. \citealt{2020A&A...642A.191E}). We find that in our case, both terms on the r.h.s of \autoref{eq:fisher} have non-negligible contribution, consistent with the findings in \cite{2024arXiv240112277P}.

The total data covariance matrix is the sum of signal covariance and noise covariance.  {As discussed in \hyperref[subsec:nonuniform]{Section \ref{subsec:nonuniform}}, the signal covariance is calculated with the baseline distribution propagated into the power spectra, and the full covariance, including cross-correlation between different wedges as shown in \autoref{fig:corr_full}, is used to calculate the Fisher matrix.} As mentioned in \hyperref[sec:psmulti]{Section \ref{sec:psmulti}}, we use the quadratic estimator formalism to calculate the noise covariance, which is then added to the signal covariance. The amplitude of the thermal noise fluctuation is calculated assuming 120\,hrs of integration time following \autoref{eq:sigman} as discussed in \hyperref[subsec:obssim]{Section \ref{subsec:obssim}}.

\subsection{Detectability of multipoles}
\label{subsec:detect}
Using the total data covariance, we present the forecasts on the measurement errors of power spectrum multipoles for SKA-Low in \autoref{fig:forecast}. To examine which scales can be probed by SKA-Low, we calculate a qualitative upper limit shown as the black dashed lines, above which the measurements provide negligible information. The limits are calculated using the maximum and the minimum of the multipoles in different wedges shown in \autoref{fig:forecast} and correspond to $\Delta^{0,2,4}_{\rm limit} = 10, 50, 50\,{\rm mK^2}$. For reference, the forecasts for measurement errors for the spherically averaged power spectra are also shown in \autoref{fig:forecast}.

For 120\,hrs of integration time, SKA-Low will be able to measure the multipoles up to $k\sim 0.2 \,{\rm Mpc^{-1}}$. Accurately modelling the data covariance is therefore essential, as the signal at small scales is non-Gaussian. Since the baseline distribution peaks around $k_\perp \sim 0.1 \,{\rm Mpc^{-1}}$, sensitivities at smaller scales decrease sharply, and there is limited amount of information that can be extracted. For a complete, futuristic EoR survey, measurements down to $k\sim 1.0 \,{\rm Mpc^{-1}}$ can be reached for $\sim 10,000$\,hrs of integration time. We refrain from discussing such a scenario. As discussed later in \hyperref[sec:values]{Section \ref{sec:values}}, we find that per-cent level precision can be reached for $\sim 120\,$hrs. For accurate forecasts of $\sim 10,000$\,hrs, the requirements on the accurate modelling of the signal and its covariance will be very stringent, and are beyond the scope of this work.

\section{Results}
\label{sec:values}
Using the calculated data covariance and the Fisher matrix formalism, we present the forecasts for constraints on EoR parameters for SKA-Low. In order to compare using multipoles as summary statistics with other approaches, we consider four scenarios: Using only the spherically averaged monopole (``mono+avg''), the spherically averaged multipoles (``multi+avg''), the monopole in clustering wedges (``mono+wedge'') and the multipoles in clustering wedges (``multi+wedge''). Both the faint and the bright models are considered, in order to demonstrate the constraining power for different types of reionization morphology.

\begin{table*}
    \centering
    \begin{tabular}{|c|c|c|c|c|c|c|c|c|c|c|}
    \hline
    model & \multicolumn{5}{c}{faint} &  \multicolumn{5}{c}{bright} \\
    \hline
    parameter & fid & mono+avg & multi+avg & mono+wedge & multi+wedge & fid & mono+avg & multi+avg & mono+wedge & multi+wedge\\
    \hline
    $(\sigma)\xi$  & 65 & 8.145 & 3.823 & 4.310 & 2.873 & 150 & 2.624 & 1.229 & 1.509 & 0.839 \\
    $(\sigma){\rm log}_{10}[T_{\rm vir}/{\rm K}]$  & 4.7 & 0.064 & 0.031 & 0.036 & 0.024 & 5.1 & 0.033 & 0.017 & 0.019 & 0.013 \\
    $\rho_{\zeta {\rm log}T_{\rm vir}}$  & / & 0.953 & 0.890 & 0.913 & 0.879 & / & 0.110 & -0.096 & -0.036 & -0.149 \\
    ${\rm det}[\mathcal{F}]$  & / & 40.41 & 335.20 & 256.19 & 958.42 & / & 138.66 & 2249.04 & 1217.63 & 8438.86 \\
    \hline
    \end{tabular}
    \caption{Forecasts for the constraining power of 21\,cm surveys using SKA-Low for EoR model inference. Two parameters, the ionization efficiency $\xi$ and the minimum virial temperature ${\rm log}_{10}[T_{\rm vir}/{\rm K}]$, are considered here. The correlation coefficient of the two parameters, $\rho_{\zeta {\rm log}T_{\rm vir}}$, and the determinant of the Fisher matrix, ${\rm det}[\mathcal{F}]$, are also shown for reference. The faint and bright models are as defined in \hyperref[subsec:21cmsim]{Section \ref{subsec:21cmsim}}. From left to right, each column shows the fiducial value, the forecasts for using only the spherically averaged monopole (``mono+avg''), the spherically averaged multipoles (``multi+avg''), the monopole in clustering wedges (``mono+wedge'') and the multipoles in clustering wedges (``multi+wedge''). }
    \label{tab:pars}
\end{table*}

In \autoref{tab:pars}, we list the forecasts for the $1\sigma$ measurement errors of the EoR parameters. Apart from the $1\sigma$ confidence interval, we also show the correlation coefficient of the two parameters as a reference for parameter degeneracy, as well as the Fisher information volume. For both models, we can see that the constraining power of SKA-Low increases significantly by including higher-order multipoles and using clustering wedges. The measurement errors are lower by a factor of $\sim 2$ when either multipoles or clustering wedges are included. The improvement suggests that a significant amount of information lies in the anisotropy of the 21\,cm power spectrum. Combining the power spectrum multipoles with clustering wedges, we find that the measurement errors decrease by a factor of $\sim 3$ comparing to using spherically averaged monopole. For a survey of $120\,$hrs, the SKA-Low will be able to constrain the ionization efficiency $\xi$ up to $\sim 3\%$ precision and the minimum virial temperature ${\rm log}[T_{\rm vir}]$ up to $\lesssim 1\%$ precision. 

\begin{figure}
    \centering
    \includegraphics[width=\linewidth]{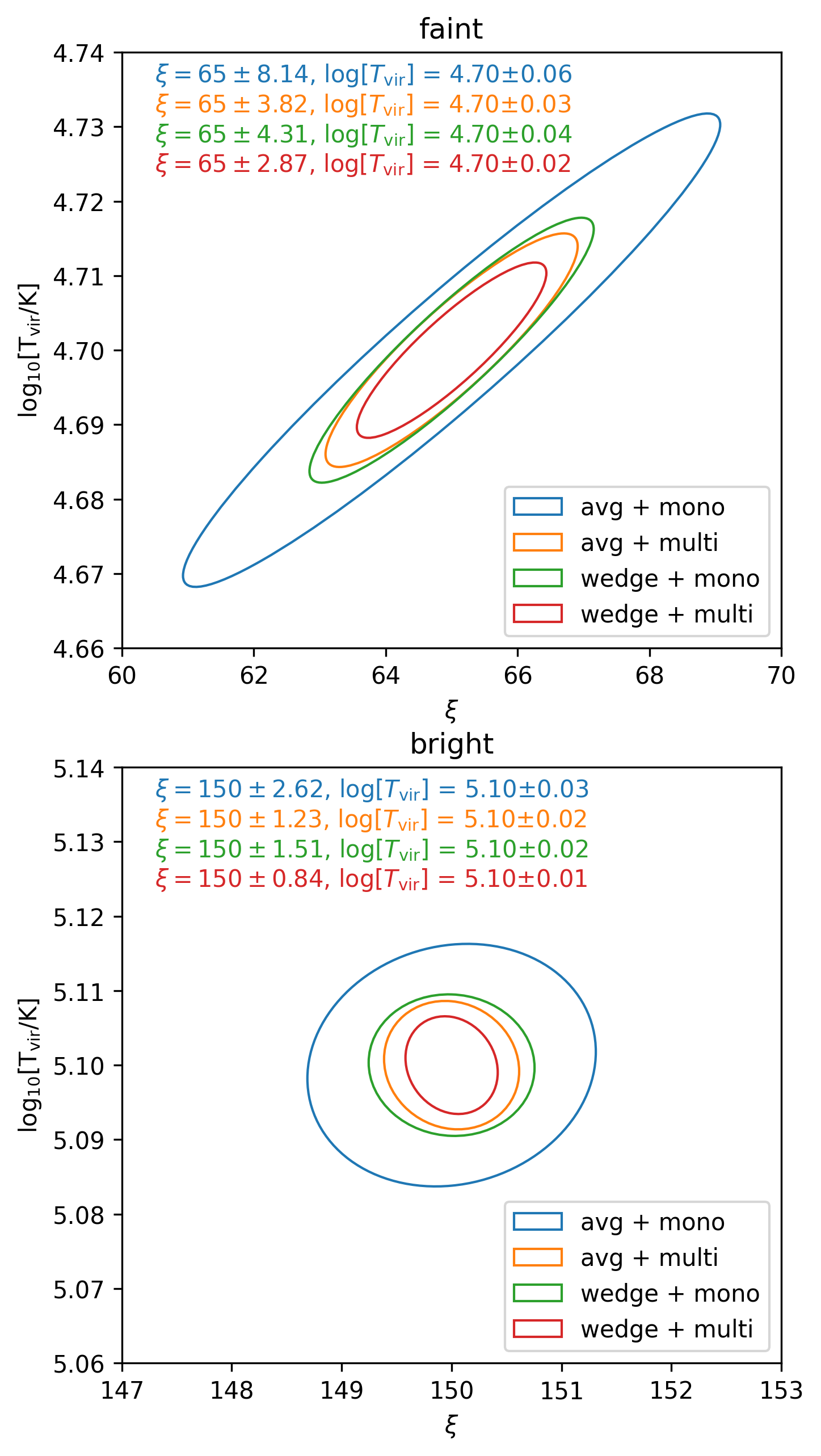}
    \caption{
    The forecasts on the $1\sigma$ confidence region of the parameter constraints from SKA-Low for the faint (top panel) and bright (bottom panel) model. Different summary statistics discussed in \hyperref[sec:values]{Section \ref{sec:values}} are colour-coded and the resulting confidence intervals are listed in the top left corner of each panel. 
    }
    \label{fig:contour}
\end{figure}

For the faint model, we find that the parameters are highly degenerate, as illustrated in \autoref{fig:div_illus}. Including the multipoles yields a slight improvement in resolving the degeneracy. The lack of improvement can be explained by the lack of information on small scales. The measurements on scales $k \gtrsim 0.1 \,{\rm Mpc^{-1}}$ are expected to probe small scales along line-of-sight, which provides information on the evolution of EoR. For the noise level assumed in this work, the small scales cannot be measured with high precision and thus the degeneracy persists. For the bright model, we find no significant correlation between the parameters. We visualise the $1\sigma$ confidence region of our forecasts in \autoref{fig:contour}.

The constraints on the EoR parameters are much more stringent for the bright model. This is due to the fact that the bright model adopts extreme fiducial values for the model parameters. As reionization is driven by massive and bright sources in the bright model, the variation in the model parameters produces bigger changes in the power spectra comparing to the faint model.

In conclusion, we find that for SKA-Low, a 21\,cm survey with $120\,$hrs of integration time will be able to measure the 21\,cm power spectrum with high sensitivity. The visibilities can be summarised into power spectrum multipoles in clustering wedges, and can be used to accurately constrain the EoR parameters.

\section{Discussions}
\label{sec:discussion}
 {We have shown in \hyperref[sec:values]{Section \ref{sec:values}} that the multipoles in clustering wedges can serve as a robust summary statistics for EoR inference. It is nevertheless important to point out that the forecasts may be subject to the specific choices of simulation configurations in this paper, which we discuss in this section.}

 {As we have discussed in this work, the improvement of parameter constraints from the power spectrum multipoles comparing to the monopole originates from the anisotropy in the power spectrum. We choose the redshift range $z\sim 7.1-8.8$ towards the end of the reionization, to maximise the effects of bubble morphology and consequently the anisotropy. For the same reason, we also do not split the frequency range into multiple sub-bands, so that the evolution along the line-of-sight is maximised. While it is beyond the scope of this work, we note that for the lower frequency range $50-145\,$MHz, the power spectrum traces more closely the underlying density and the spin temperature field. As a result, we do not expect the same level of improvements in the constraints on reionization parameters from the multipoles in clustering wedges.}

 {We also comment on the fact that the multipoles in clustering wedges can only be used when wide-field effects in EoR observations are sufficiently mitigated. Throughout this work, we have assumed the foreground wedge to be $\mu < 0.27$, corresponding to the size of the instrument beam. This leaves a large part of the $\bm{k}$-space for 21\,cm measurements, and allows us to perform multipole expansions as well as splitting the $k$-space into clustering wedges. If the physical horizon is used instead, i.e. the foreground leakage is significant for the whole sky beyond the beam field-of-view, the 21\,cm power spectrum can only be measured at $\mu>0.96$. In that case, the multipole expansions and clustering wedges will not be meaningful since there is almost no variation for $\mu$. We find that for the spherically averaged monopole, the measurement errors are enlarged by a factor of $\sim 2.5$ when the foreground wedge is at the physical horizon, as shown in \hyperref[apdx:horizon]{Appendix \ref{apdx:horizon}}.}

 {We note that the degeneracy of the parameters is discussed in the simplified setting of the EoR simulation, where only two parameters $\xi$ and $T_{\rm vir}$ are considered. For simulations with an extended set of model parameters, the degeneracies between parameters, and how multipoles in clustering wedges may be used to resolve them, can be more complicated. As we have shown in \hyperref[apdx:convergence]{Appendix \ref{apdx:convergence}}, the covariance calculation converges when averaged across 10 realizations. For a more realistic reionization model, the number of realizations needed is likely to increase. The noise covariance is calculated analytically as described in \hyperref[apdx:est]{Appendix \ref{apdx:est}}, leading to a total covariance matrix that is numerically stable when inverted. For a more optimistic assumption of the total observation time $t_{\rm tot} \gg 120\,$hrs, the signal covariance will be dominant and demand a large number of realizations for accurate modelling.  {Accurate description of the degeneracies between a large number of parameters requires Bayesian inference of the posterior instead of the Fisher matrix analysis, which we leave for future work.}
}

 {Finally, we comment on the relation between the multipoles in clustering wedges and the cylindrical power spectrum. The multipoles and clustering wedges can be viewed as alternative ways of binning the $\bm{k}$-space, as opposed to binning in terms of $\bm{k}_\perp$ and $k_\parallel$. In theory, the cylindrical power spectrum should have the complete information content of the two-point statistics of the EoR signal. However, the cylindrical binning of $\bm{k}$-space results in a signal covariance matrix that is hard to model. The fine resolution in $\bm{k}$-space requires a large number of realizations for each parameter set for convergence. For example, \cite{2024arXiv240112277P} uses 130 realizations for each sample of the parameter space. Moreover, many observational effects will introduce mode-mixing to the power spectrum, such as primary beam, bandpass tapering and chromatic sampling rate. All of these effects further complicate the calculation of the signal covariance, which are beyond the scope of this work to quantify. The multipoles, on the other hand, are calculated in 1D $k$-bins, and therefore are less susceptible to these effects.}

\section{Conclusions}
\label{sec:conclusion}
In this paper, we explore the advantages of using power spectrum multipoles and clustering wedges as summary statistics over using the spherically averaged monopole power spectrum for EoR measurements. In particular, we comprehensively demonstrate where additional information arises in the power spectrum multipoles and in the partition of the cylindrical $\bm{k}$-space. We then present forecasts for parameter constraints using multipole power spectra and clustering wedges for SKA-Low observations of redshift range $z\sim 7.1-8.8$.

We confirm the importance of extracting information from power spectrum anisotropy. Through simulating mock 21\,cm observations for future SKA-Low surveys using different model parameters for the EoR, we verify that the 1D monopole power spectrum is not optimal in distinguishing physical models of reionization, as it averages over the entire $\bm{k}$-space without retaining the anisotropic information of the 21\,cm signal. We point out that power spectrum multipoles provide a direct summary of the anisotropic 21\,cm cylindrical power spectrum. There is visible improvement in the constraining power from the amplitude of the power spectra when multipoles are included. Small scale measurements of multipoles probe small-scale evolution along the line-of-sight and contain information for constraining EoR parameters.

EoR signals at small scales are dominated by the morphology of the ionization bubbles. The resulting monopole power spectrum is therefore highly correlated, especially for the latter stage of reionization considered in this work. We quantify the covariance of the signal with a large suite of simulations of the reionization lightcone, and find that small scales $k\gtrsim 0.2\,{\rm Mpc^{-1}}$ are almost completely correlated for the monopole. We verify that this correlation is indeed caused by the growth of ionization bubbles, as simulations with larger and fewer bubbles show stronger correlation at large scales. To our knowledge, we are the first to point out that this correlation can be disentangled by including the quadrupole and hexadecapole power spectra. The higher-order multipoles probe line-of-sight scales along which reionization evolve rapidly, providing information beyond the compressed 1D fluctuation. 

The baseline distribution of a radio interferometer is not uniform. As a result, we find that for SKA-Low, the sampling in the cylindrical $\bm{k}$-space is skewed heavily towards large values of $\mu$ for scales $k\gtrsim 0.2\,{\rm Mpc^{-1}}$ when averaged into 1D power spectra. Therefore, measurements at large $k$ effectively only probe a small region of $\bm{k}$-space with high $k_\parallel$. This calls for the partitioning of $\bm{k}$-space into clustering wedges. Within each clustering wedge, the sampling of $\mu$ becomes relatively uniform. The power spectrum multipoles can then be used to measure the $k_\parallel$-dependence of the signal, optimally extracting information on the EoR. Accurate calculation of the signal covariance will be crucial for model inference, as there are significant contributions from higher-order correlations in the covariance matrix.

In order to estimate the constraining power of the multipoles and clustering wedges for the EoR, we perform Fisher matrix forecasts to quantify the error covariance of reionization parameters assuming $120\,$hrs of observation with SKA-Low. We find that on large scales $k\lesssim 0.1\,{\rm Mpc^{-1}}$, the multipole power spectra can be measured with relatively high precision. As a result, SKA-Low will be able to give stringent constraints on EoR parameters, with the measurement errors reaching $\lesssim 1\%$. Comparing to only using the spherically averaged monopole, the power spectrum multipoles in clustering wedges yield a factor of $\sim$3 improvement on parameter constraints. 

 {It is worth noting that the strong correlation between the EoR parameters does not disappear when multipoles and clustering wedges are used. We find that for small variations of the EoR parameters, the differences in the multipoles are mostly in relatively small scales.} Due to the lack of sensitivity for scales $k\gtrsim 0.1\,{\rm Mpc^{-1}}$, the degeneracy between EoR parameters is only slightly resolved by including the multipoles.  {This is due to the relatively sparse distribution of long baselines in SKA-Low, and our conservative assumption of $120\,$hrs of observation time.} For a complete 21\,cm survey with $\sim 10,000\,$hrs of integration time, SKA-Low will be able to probe into the small scales $k\sim 1\,{\rm Mpc^{-1}}$ where the multipoles can be used to further break the degeneracy.

Our work provides strong incentives for using power spectrum multipoles in clustering wedges in forthcoming EoR measurements. In addition to the advantages mentioned above, the multipoles are less susceptible to various observational systematic effects. Quantifying the covariance of the multipoles in different wedges from visibility data and using a conventional model inference framework with the multipoles are within the reach of the anticipated data quality in the near future. Robust model inference with the power spectrum multipoles and clustering wedges will provide crucial insights into the physics driving the cosmic reionization.

\section*{Acknowledgements}
ZC and AP are funded by a UKRI Future Leaders Fellowship [grant MR/X005399/1; PI: Alkistis Pourtsidou]. The computation underlying this work is performed on the \textit{cuillin} cluster located at the Institute for Astronomy, University of Edinburgh. Apart from aforementioned packages, this work also uses \textsc{pytorch} \citep{NEURIPS2019_9015}, \textsc{numpy} \citep{2020Natur.585..357H}, \textsc{scipy} \citep{2020NatMe..17..261V}, \textsc{astropy} \citep{2018AJ....156..123A}, \textsc{camb} \citep{2000ApJ...538..473L}, \textsc{CASA} \citep{2022PASP..134k4501C} and \textsc{matplotlib} \citep{Hunter:2007}. 

For the purpose of open access, the author has applied a Creative Commons Attribution (CC BY) licence to any Author Accepted Manuscript version arising from this submission.

\section*{Data Availability}
The scripts for simulating reionization lightcones are publicly available as a GitHub repository\footnote{\url{https://github.com/zhaotingchen/eor_fisher_pipeline}}.
The scripts for visibility simulation and the subsequent Fisher matrix analysis will be shared on reasonable request to the corresponding author.


\bibliographystyle{mnras}
\bibliography{example}



\appendix

\section{Quadratic Estimator for Power Spectrum Multipoles}
\label{apdx:est}
In this section, we discuss the quadratic estimator formalism used in this paper for power spectrum multipoles. The quadratic estimator for cosmological power spectra is well studied in the context of CMB (e.g. \citealt{1997PhRvD..55.5895T,2001PhRvD..64f3001T}), galaxy clustering (e.g. \citealt{1998ApJ...499..555T}) and more recently in interferometric 21\,cm observations (e.g. \citealt{2011PhRvD..83j3006L,2021MNRAS.501.1463K}). For this work, we follow the formalism in \cite{2021MNRAS.501.1463K} and \cite{2023MNRAS.518.2971C}, and specify our estimator in the context of the delay power spectrum from visibilities. We then generalise it to power spectrum multipoles.

A radio interferometer measures the sky emission with baselines, i.e. pairs of antennas, as visibilities \citep{2011A&A...527A.106S}
\begin{equation}
\begin{split}
    V(u,v,w,f) = \iint & \frac{{\rm d}l\,{\rm d}m}{\sqrt{1-l^2-m^2}}\, I(l,m,f)\,A(l,m,f)\\
    \times&\,{\rm exp}\big[ -2\pi(lu+mv+\sqrt{1-l^2-m^2}w)\big],
\end{split}
\label{eq:rime}
\end{equation}
where $I(l,m,f)$ is the flux density of the sky signal at frequency $f$ and angular coordinate on the sky $(l,m)$, and $A(l,m,f)$ is the power response of the instrument. For a pair of antennas, the distance vector between them, denoted as $\bm{b}$, can be written in East-North-Up (ENU) coordinate frame as 
\begin{equation}
    \bm{b} = \lambda(u,v,w)^{\rm T},
\end{equation}
where $\rm T$ denotes the transpose operation and $\lambda$ is the observing wavelength. 

For each instantaneous baseline, the visibilities across the frequency channels can be Fourier transformed so that
\begin{equation}
    \tilde{V}(u,v,\eta) = \int {\rm d}f\, V(u,v,f) \,{\rm exp}\big[-2\pi\eta f\big].
\end{equation}
For the 21\,cm signal, the coordinates $(u,v)$ reflect the angular scale in Fourier space while the delay time $\eta$ reflects the Fourier mode of density fluctuation along the line-of-sight:
\begin{align}
    \bm{k}_\perp = \frac{2\pi \bm{u}}{X}\\
    k_\parallel = \frac{2\pi\eta}{Y},
\end{align}
where $\bm{u} = (u,v)^{\rm T}$.

For a bandpower in the $\alpha^{\rm th}$ $\bm{k}$-bin of a multipole moment $\ell$, which we denote as $p^\ell_\alpha$, the quadratic estimator can be written as 
\begin{equation}
    \hat{p}^\ell_\alpha = C_{\rm \hi}\, \big(\mathbf{V}^\dagger \, \mathbf{E}_\alpha^\ell \,\mathbf{V} - \hat{b}^\ell_\alpha\big),
\label{eq:est}
\end{equation}
where $\mathbf{V}$ is the visibility data vector, $\dagger$ denotes the Hermitian transpose, $\mathbf{E}_\alpha^\ell$ is an estimation matrix discussed later and $\hat{b}^\ell_\alpha$ is the bias correction term. $C_{\rm \hi}$ is a factor that converts the flux density to brightness temperature unit and renormalises the survey volume \citep{2014ApJ...788..106P}
\begin{equation}
    C_{\rm \hi} = \bigg(\frac{\lambda^2}{2k_{\rm B}}\bigg)^2 \frac{X^2Y}{\Omega_{\rm ps}B},
\end{equation}
where $\lambda$ is the observing frequency, $k_{\rm B}$ is the Boltzmann constant, $X$ is the comoving distance of the 21\,cm signal at the observing frequency, $B$ is the frequency bandwidth of the observation, and $\Omega_{\rm ps}$ is the power-squared beam area defined as 
\begin{equation}
    \Omega_{\rm ps} = \int {\rm d}l{\rm d}m |A(l,m)|^2.
\end{equation}
$Y$ is the comoving length scale per frequency interval
\begin{equation}
    Y = \frac{\lambda (1+z)}{H(z)},
\end{equation}
where $z$ is the redshift of the 21\,cm signal at the observing frequency, and $H(z)$ is the Hubble parameter.

The data vector comprises visibilities measured at each baseline at each time step in each frequency channel. For an ideal experiment where the data vector consists only of \hi\ signal and the beam response is unitary, the 21\,cm power spectrum for $\bm{k}_\alpha$ can be written as \citep{2006ApJ...653..815M,2012ApJ...756..165P,2014ApJ...788..106P}
\begin{equation}
    P_{\rm 21}(\bm{k}_\alpha) = C_{\rm \hi}|\tilde{V}^{\rm ideal}(u_\alpha,v_\alpha,\eta_\alpha)|^2.
\label{eq:psideal}
\end{equation}

The power beam of the instrument is multiplied with the signal on the sky, and therefore convolves the signal in Fourier space. The mode-mixing can be written in the data vector as
\begin{equation}
    \tilde{V}^\ell_\alpha = \sum_\beta \mathcal{M}^\ell_{\alpha\beta}\tilde{V}^{\ell ,\rm ideal}_{\beta}, \, \tilde{V}^\ell_\alpha = \tilde{V}^\ell(\bm{k}_\alpha),
\end{equation}
where $\mathcal{M}^\ell_{\alpha\beta}$ is the mode-mixing matrix that can be calculated given the power beam, and $\tilde{V}^\ell_\alpha$ is the delay-transformed data vector for multipole $\ell$:
\begin{equation}
    \tilde{\mathbf{V}}^\ell = \mathcal{F}_\ell \, \mathbf{V}.
\end{equation}
$\mathcal{F}_\ell$ is the Fourier transformation kernel that transforms the visibility to multipole overdensity. In this paper, we consider $\ell = (0,2,4)$. $\mathcal{F}_0$ is simply the Discrete Fourier Transform kernel. For higher moments, in the flat-sky and plane-parallel limit we have \citep{2015PhRvD..92h3532S}
\begin{equation}
    \mathcal{F}_2 = \frac{3}{2}{\rm diag}\big[ \bm{\hat{k}}_\parallel^2 \big] \mathcal{F}_0-\frac{1}{2}\mathcal{F}_0,
\end{equation}
\begin{equation}
    \mathcal{F}_4 = \frac{35}{8}{\rm diag}\big[ \bm{\hat{k}}_\parallel^4 \big]\mathcal{F}_0 - \frac{5}{2}\mathcal{F}_2-\frac{7}{8}\mathcal{F}_0,
\end{equation}
where ${\rm diag}[]$ denotes the operation that diagonalises a vector into a matrix, $\bm{\hat{k}}_\parallel$ is a vector so that $\bm{\hat{k}}_\parallel^\alpha = \bm{\mu}^\alpha = k_\parallel^\alpha / |\bm{k}_\alpha|$.

Similar to \autoref{eq:psideal}, we can then write down the estimator when the measurement is ideal:
\begin{equation}
    \hat{p}^{\rm ideal,\ell}_\alpha = C_{\rm \hi}\big(\tilde{V}^{\ell,\rm ideal}\big)^\dagger \,{\rm diag}\big[\bm{w}_\alpha\big] \,\tilde{V}^{\rm 0, ideal},
\end{equation}
where $\bm{w}_\alpha$ is a selection vector with the $\alpha^{\rm th}$ element being 1 and all other elements 0.

We can then factorize the estimator of \autoref{eq:est} in a way that relates to the Fourier multipole overdensity
\begin{equation}
\begin{split}
    \langle \hat{p}^\ell_\alpha \rangle =& C_{\rm \hi}\bigg\langle \big(\tilde{\mathbf{V}}^{\ell}\big)^\dagger \,\big(\mathcal{F}_\ell^{-1}\big)^\dagger \, \mathbf{E}^\ell_\alpha \mathcal{F}_0^{-1} \, \tilde{\mathbf{V}}^{0} \bigg\rangle \\
    =& C_{\rm \hi}\bigg\langle \big(\tilde{\mathbf{V}}^{\ell,\rm ideal}\big)^\dagger \, \big(\mathcal{M}^\ell\big)^\dagger \, \big(\mathcal{F}_\ell^{-1}\big)^\dagger \, \mathbf{E}^\ell_\alpha \, \mathcal{F}_0^{-1} \, \mathcal{M}^0 \, \tilde{\mathbf{V}}^{0,\rm ideal} \bigg\rangle\\
    =& C_{\rm \hi}\,{\rm tr}\Big[ \tilde{\mathbf{E}}^\ell_\alpha \big\langle \tilde{\mathbf{V}}^{0,\rm ideal} \,\big(\tilde{\mathbf{V}}^{\ell,\rm ideal}\big)^\dagger \big\rangle \Big],
\end{split}
\end{equation}
where the bias term is neglected for now, ${\rm tr}[]$ denotes the trace of a matrix and we have defined a mode-mixed Fourier space estimator
\begin{equation}
    \tilde{\mathbf{E}}^\ell_\alpha = \big(\mathcal{M}^\ell\big)^\dagger \, \big(\mathcal{F}_\ell^{-1}\big)^\dagger \, \mathbf{E}^\ell_\alpha \, \mathcal{F}_0^{-1} \, \mathcal{M}^0.
\end{equation}
Assuming that the 21\,cm field does not correlate for different $\bm{k}$-modes, the covariance matrix of the ideal signal can be written as
\begin{equation}
    C_{\rm \hi}\big\langle \tilde{\mathbf{V}}^{0,\rm ideal} \,\big(\tilde{\mathbf{V}}^{\ell,\rm ideal}\big)^\dagger \big\rangle = \sum_\beta {\rm diag}\big[ \bm{w}_\beta \big] p_{\beta}^\ell,
\end{equation}
where $\beta$ loops over every measured $\bm{k}$-mode and $p_{\beta}^\ell$ is the power spectrum multipole at $\bm{k}_\beta$.

The window function of the estimator is then
\begin{align}
    \langle \hat{p}^\ell_\alpha \rangle =& \sum_\beta \mathbf{W}_{\alpha\beta}^\ell p^\ell_\beta  = \sum_\beta {\rm tr}\Big[ \tilde{\mathbf{E}}^\ell_\alpha \, {\rm diag}\big[ \bm{w}_\beta \big]\Big]p_{\beta}^\ell,\\
    \mathbf{W}_{\alpha\beta}^\ell =& \,{\rm tr}\Big[ \tilde{\mathbf{E}}^\ell_\alpha \, {\rm diag}\big[ \bm{w}_\beta \big]\Big].
    \label{eq:window}
\end{align}

To remove the bias from noise and foregrounds in the estimator, simply note $\langle \hat{p}^\ell_\alpha \rangle = C_{\rm \hi}\,{\rm tr}[\mathbf{E}^\ell_\alpha \mathbf{C} ]$ where $\mathbf{C}$ is the data covariance and the bias correction term is therefore
\begin{equation}
    \hat{b}^\ell_\alpha = C_{\rm \hi}\,{\rm tr}[\mathbf{E}^\ell_\alpha (\mathbf{C}_{\rm N}+\mathbf{C}_{\rm fg}) ],
\end{equation}
where $\mathbf{C}_{\rm N}$ and $\mathbf{C}_{\rm fg}$ are the noise and foreground covariance, respectively. Throughout this paper, we adopt the foreground avoidance method \citep{2012ApJ...752..137M} and only estimate the power spectrum at $\bm{k}$-modes where the foreground power is negligible.

The estimator can be empirically factorized as
\begin{equation}
    \mathbf{E}^\ell_\alpha = \sum_\gamma \mathbf{S}_{\alpha\gamma}^\ell\,\mathbf{R}^\dagger \, \mathcal{F}_\ell^\dagger \, {\rm diag}\big[ \bm{w}_\gamma \big] \, \mathcal{F}_0 \, \mathbf{R},
\label{eq:estdecom}
\end{equation}
where $\mathbf{S}^\ell$ is a renormalisation matrix to be determined later, and $\mathbf{R}$ is a weighting matrix that can include various data analysis techniques such as frequency tapering, inverse noise covariance weighting, and foreground cleaning. 

Substituting \autoref{eq:estdecom} into \autoref{eq:window}, the window function becomes
\begin{equation}
\begin{split}
    \mathbf{W}_{\alpha\beta}^\ell = & \sum_\gamma \mathbf{S}_{\alpha\gamma}^\ell\,{\rm tr}\Big[ \big(\mathcal{M}^\ell\big)^\dagger \, \big(\mathcal{F}_\ell^{-1}\big)^\dagger \, \mathbf{R}^\dagger \, \mathcal{F}_\ell^\dagger \, {\rm diag}\big[ \bm{w}_\gamma \big] \\
    &\times \mathcal{F}_0 \, \mathbf{R} \, \mathcal{F}_0^{-1} \, \mathcal{M}^0 \, {\rm diag}\big[ \bm{w}_\beta \big]\Big]\\
    = & \sum_\gamma \mathbf{S}_{\alpha\gamma}^\ell\, {\rm tr}\Big[ \big(\mathcal{M}^\ell\big)^\dagger \, \tilde{\mathbf{R}}_\ell^\dagger \, {\rm diag}\big[ \bm{w}_\gamma \big] \, \tilde{\mathbf{R}}_0^\dagger \, \mathcal{M}^0 \, {\rm diag}\big[ \bm{w}_\beta \big] \Big].
\end{split}
\end{equation}
Therefore, we can form the quantity
\begin{equation}
    \mathbf{H}_{\alpha\beta}^\ell = {\rm tr}\Big[ \big(\mathcal{M}^\ell\big)^\dagger \, \tilde{\mathbf{R}}_\ell^\dagger \, \tilde{\mathbf{R}}_0^\dagger \, \mathcal{M}^0 \, {\rm diag}\big[ \bm{w}_\alpha \big]  \, {\rm diag}\big[ \bm{w}_\beta \big] \Big],
\end{equation}
so that
\begin{equation}
    \mathbf{W}^\ell = \mathbf{S}^\ell \,\mathbf{H}^\ell,
\label{eq:wsh}
\end{equation}
where we utilise the fact that ${\rm diag}[ \bm{w}_\alpha ]$ is commutable with any matrix.

The formulation of the window function in \autoref{eq:wsh} is determined by the choice of $\mathbf{R}$, which then determines $\mathbf{H}^\ell$, and the subsequent choice of $\mathbf{S}^\ell$. For example, a common choice of $\mathbf{S}^\ell$ is the diagonal matrix $\mathbf{S}^\ell_{\alpha\beta} = \delta^{\rm K}_{\alpha\beta} (\mathbf{H}^\ell_{\alpha\beta})^{\rm -1}$ (e.g. \citealt{1997MNRAS.289..285H}), where $\delta^{\rm K}$ is the Kronecker delta. $\mathbf{S}^\ell = (\mathbf{H}^\ell)^{-1}$ decorrelates the estimated bandpowers. Throughout this paper, we choose $\mathbf{S}^\ell = (\mathbf{H}^\ell)^{-1/2}$ to balance the measurement errors and correlation of different $k$-bins.

Assuming that the effects of mode-mixing are negligible in the signal covariance, the covariance matrix of the estimator can be written as
\begin{equation}
    \Sigma_{\alpha\beta}^{\ell_1 \ell_2} = \langle \hat{p}^{\ell_1}_\alpha \hat{p}^{\ell_2}_\beta \rangle \approx {\rm tr}\big[ \mathbf{C}\,\mathbf{E}^{\ell_1}_\alpha \,\mathbf{C}\,\mathbf{E}^{\ell_2}_\beta \big] + {\rm tr}\big[ \mathbf{C}_{\rm s}\,\mathbf{E}^{\ell_2}_\beta\, \mathbf{C}_{\rm s}\,\mathbf{E}^{\ell_1}_\alpha \big],
\end{equation}
where $\mathbf{C}_{\rm s}$ is the signal covariance. This is different from the case in e.g. the CMB, due to the fact that thermal noise in visibility is complex. The above equation also assumes that the $u$-$v$ coverage of the interferometer baselines is complete.

While the estimator formalism presented in this section is generic and takes into account the entire data vector, constructing the entire data covariance matrix is too computationally expensive for the scope of our work. We implement the estimator at each $u$-$v$ grid to compute the 3D band powers and covariances, and then perform the 1D binning assuming that different $u$-$v$ grids do not correlate. Since we are only using the quadratic estimator to calculate noise covariance in this work, this simplification does not impact our results. A full detection methodology from visibility data to multipoles is left for future work.

\begin{figure}
    \centering
    \includegraphics[width=\linewidth]{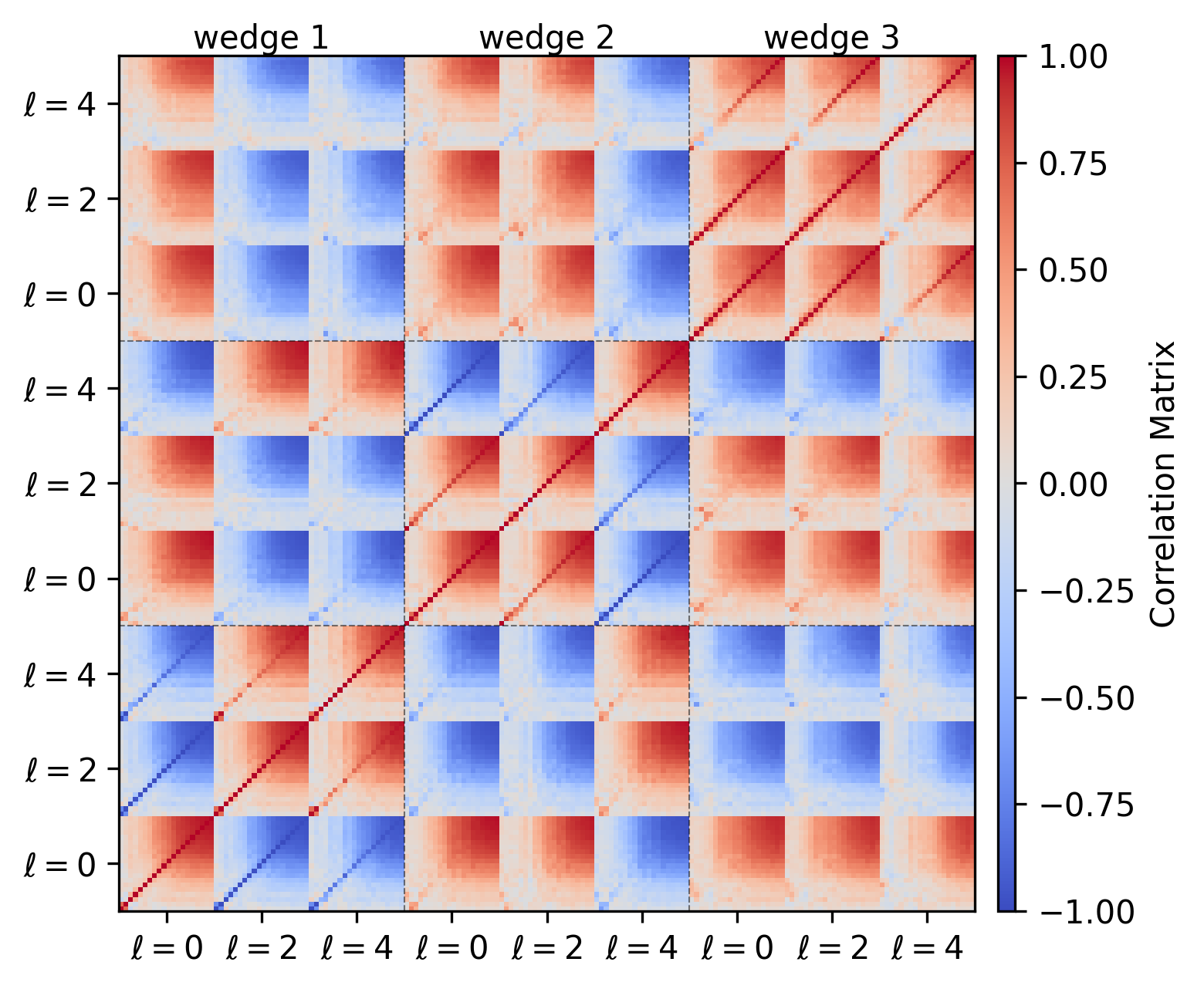}
    \caption{
    The correlation matrix of the power spectrum multipoles. The $k$ values of the axes are the same as \autoref{fig:corr_wedge} and are omitted for visual clarity.
    }
    \label{fig:corr_full}
\end{figure}

\begin{figure*}
    \centering
    \includegraphics[width=0.32\linewidth]{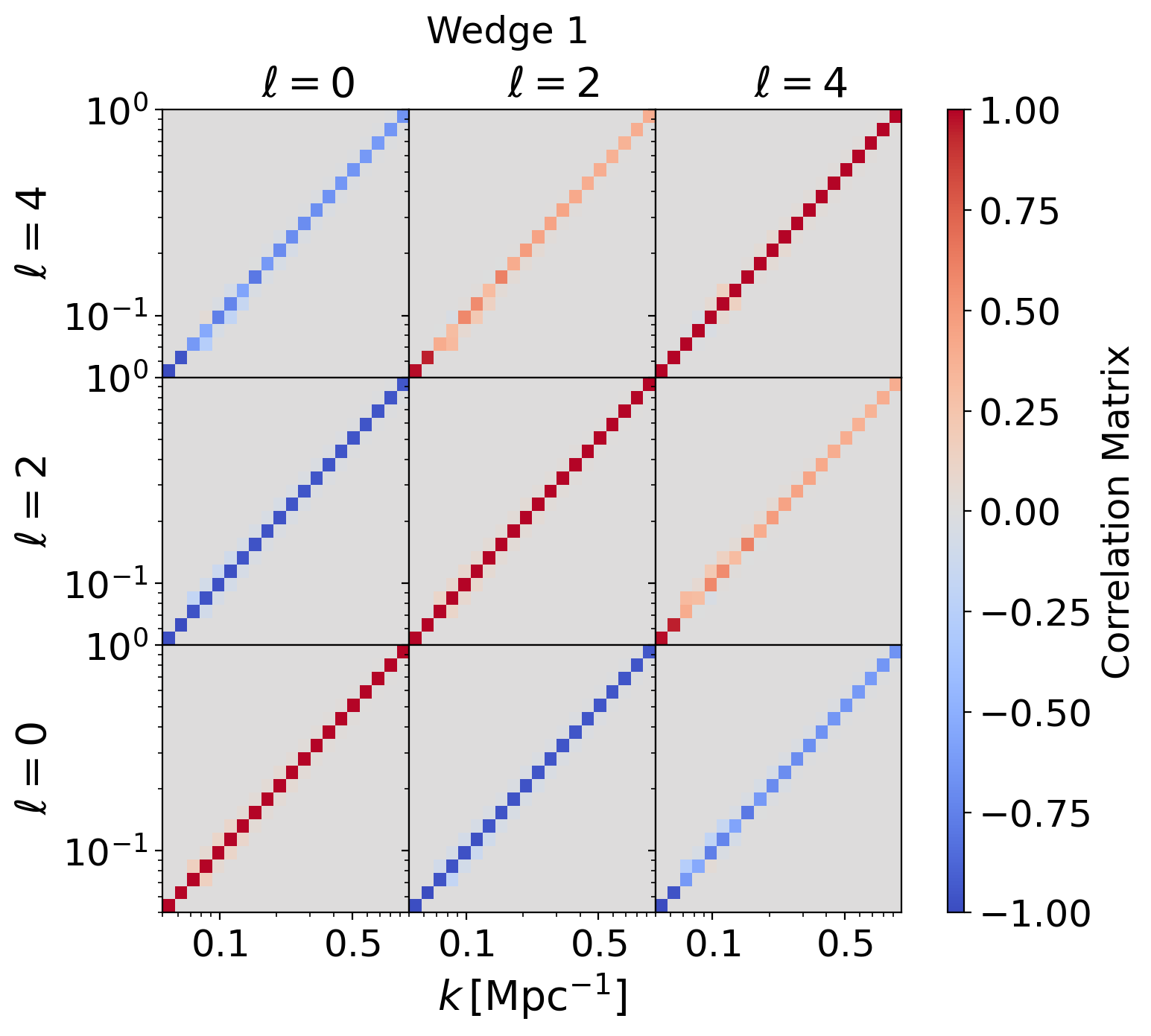}
    \includegraphics[width=0.32\linewidth]{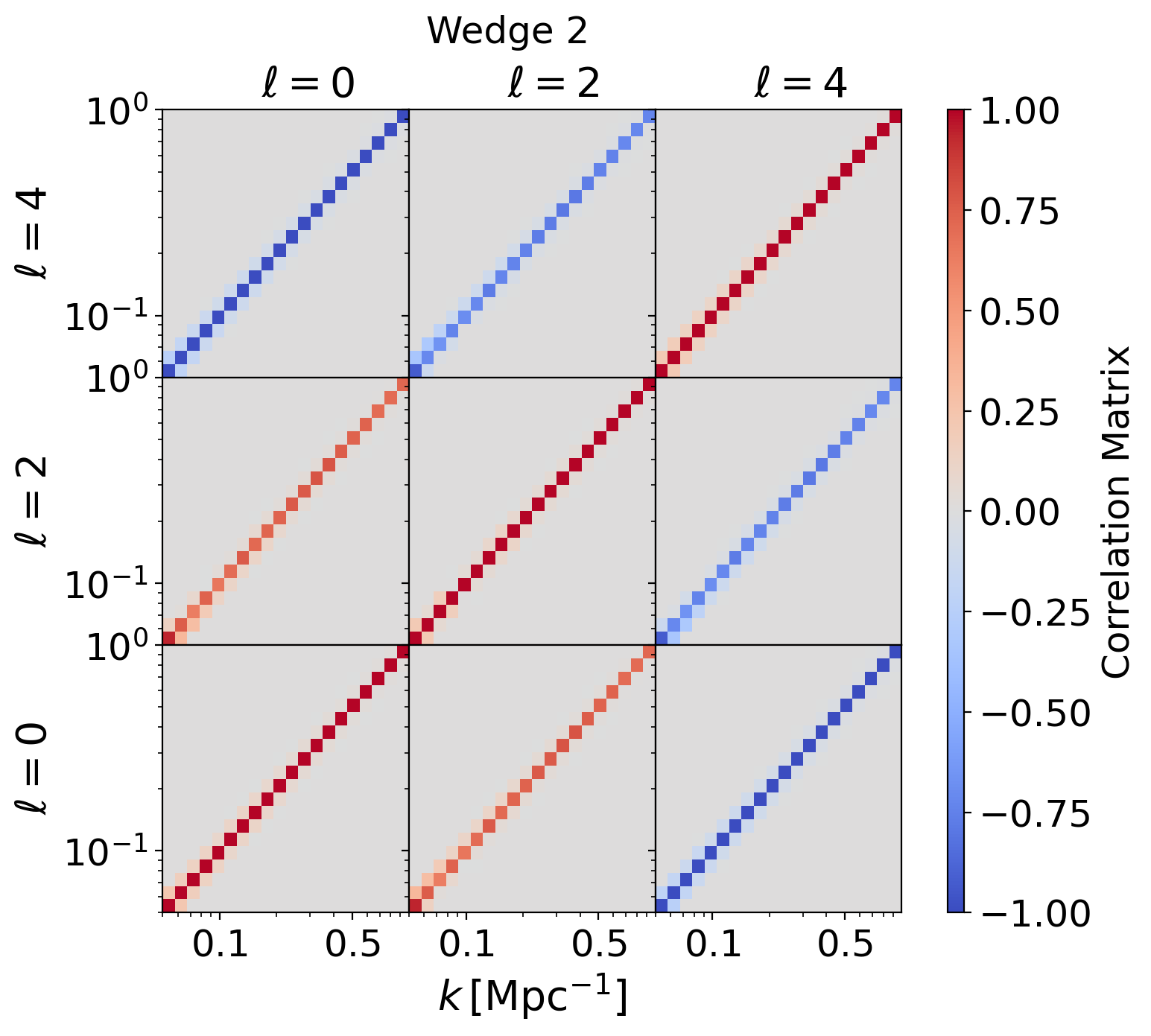}
    \includegraphics[width=0.32\linewidth]{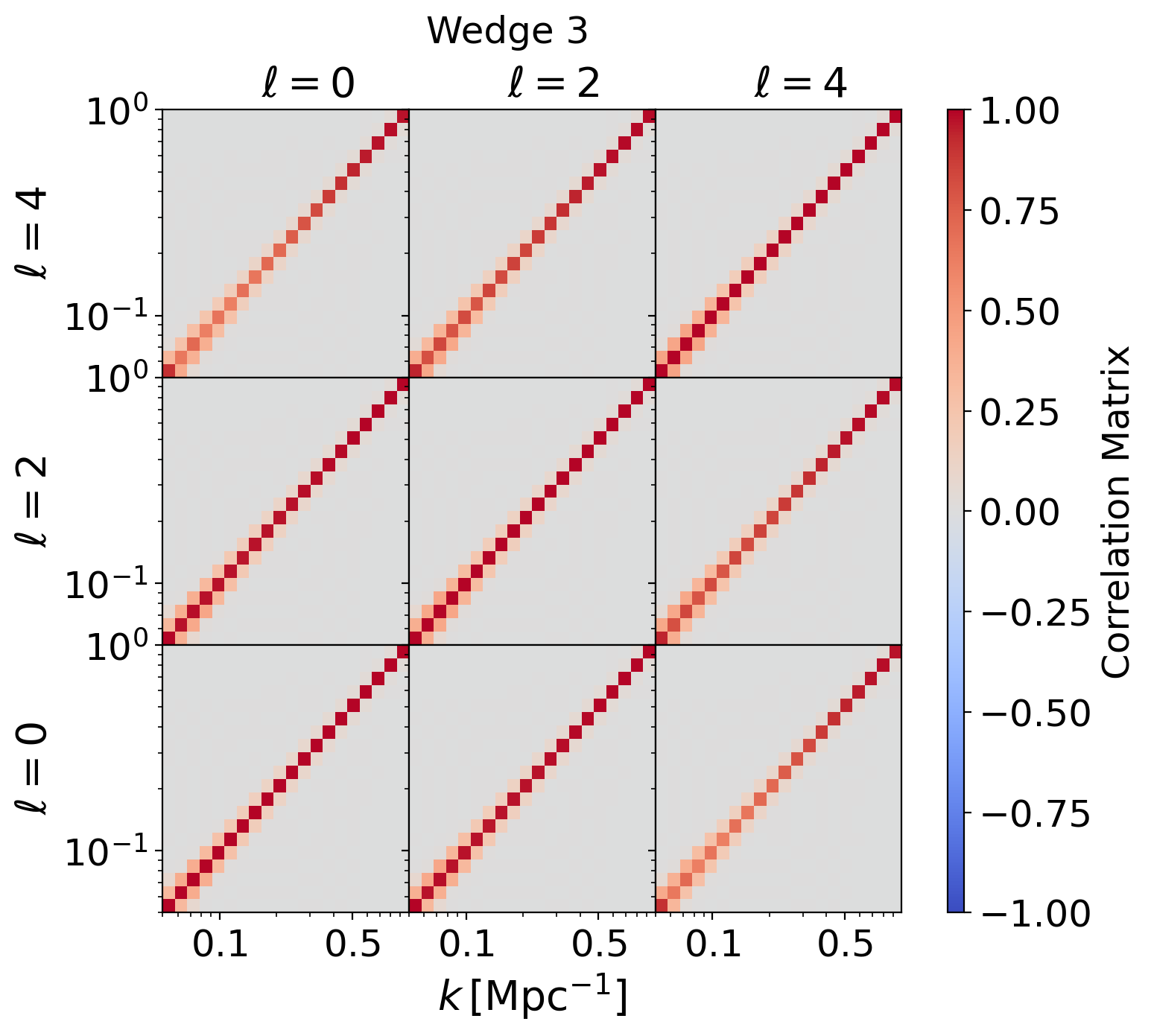}
    \caption{
    The correlation matrix of noise covariance for different clustering wedges. The amplitude of the noise covariance is shown in \autoref{fig:forecast}.
    }
    \label{fig:corr_noise}
\end{figure*}

For reference, we show the full signal correlation matrix in \autoref{fig:corr_full}, calculated using the jackknife method discussed in \hyperref[sec:psmulti]{Section \ref{sec:psmulti}}, and the full noise correlation matrix in \autoref{fig:corr_noise}, calculated using the quadratic estimator discussed above. The amplitude of the total data covariance matrices is shown in \hyperref[subsec:detect]{Section \ref{subsec:detect}}.

\section{Convergence tests}
\label{apdx:convergence}
 {In this section, we discuss the convergence of our Fisher matrix analysis. The calculation of Fisher matrix depends on the calculation of the derivatives of the summary statistics as well as the covariance with regard to the model parameters. However, the growth of ionization structures during EoR is nonlinear and non-Gaussian, leading to potential numerical instabilities in the jackknife method we use to calculate the covariance of the summary statistics. To resolve the potential instability, we average over 10 independent realizations for each set of input parameter values. We evaluate the sufficiency of the number of realizations in this section.}

 {To test the convergence of the Fisher matrix, we perform a jackknife test by excluding a certain number of realizations at a time. The rest of the realizations are then used to calculate the Fisher matrix. We then calculate the mean and the standard deviation of the measurement errors among all possible combinations. For example, to evaluate convergence of the number of realizations $N_{\rm realization} = 2$, we exclude $10 - N_{\rm realization} = 8$ realizations resulting in the total of 45 combinations.}

\begin{figure}
    \centering
    \includegraphics[width=\linewidth]{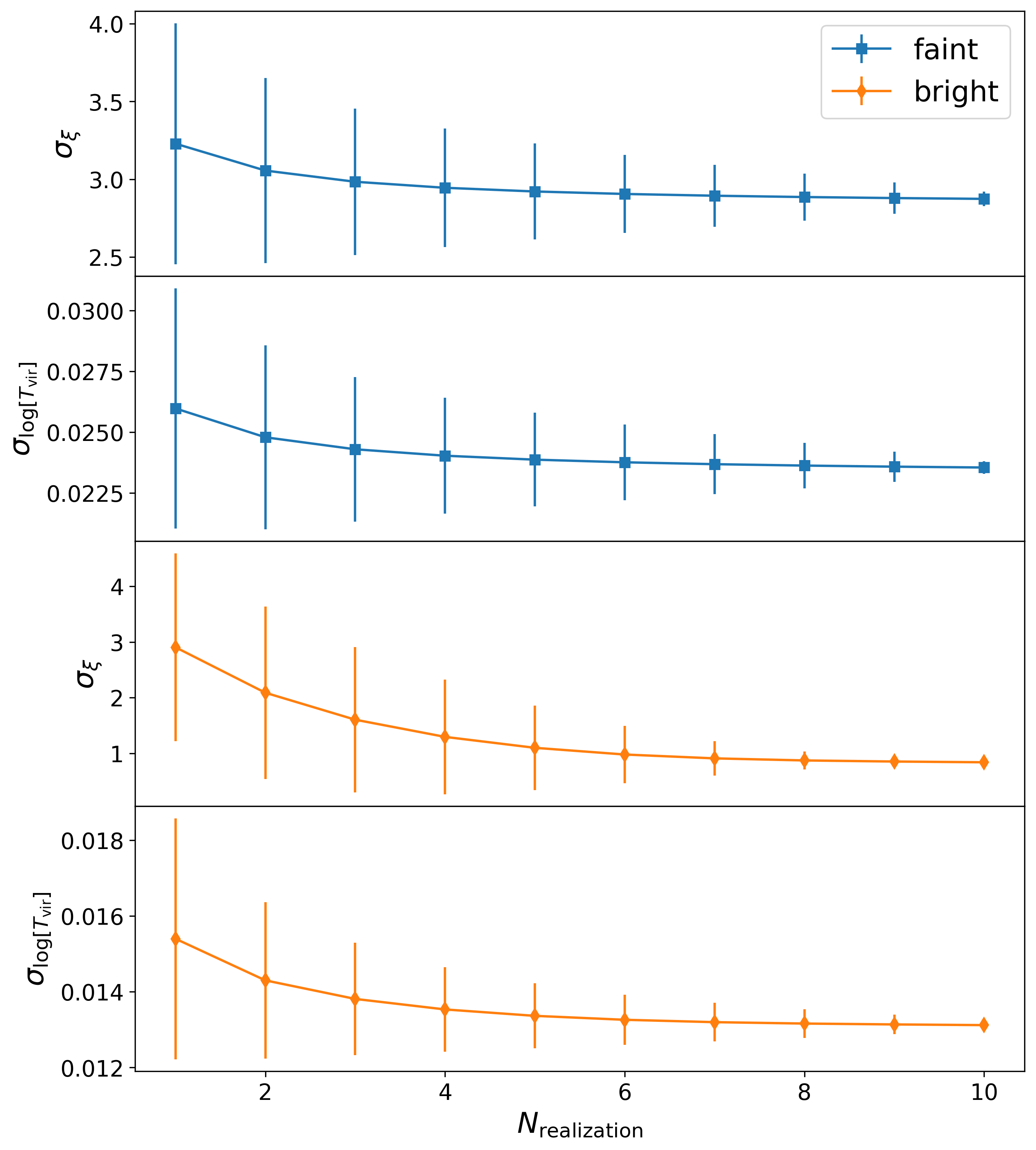}
    \caption{
    The relation between the forecasted measurement errors of EoR parameters and the number of realizations averaged. The measurement errors decrease and then plateau when number of realizations increase. The error bars of the measurement errors are the standard deviations among the possible combinations. Only results for using the multipoles in clustering wedges are shown for simplicity.
    }
    \label{fig:converge}
\end{figure}

 {The resulting relation between the number of realizations and the projected measurement errors of the parameters is shown in \autoref{fig:converge}. For simplicity, only the scenario of using the multipoles in clustering wedges is shown. The error bars of the measurement errors are linearly extrapolated from $N_{\rm realization} = 1-9$ to $N_{\rm realization} = 10$. The measurements errors first decrease then plateau with increased number of realizations. When the number of realizations is small, the measurement errors include numerical instabilities which enlarge the error bars. As shown, the values for $\sigma_{\xi}$ and $\sigma_{{\rm log}[T_{\rm vir}]}$ plateau after around 5 realizations, suggesting that the results have converged for both models.}

\begin{figure}
    \centering
    \includegraphics[width=\linewidth]{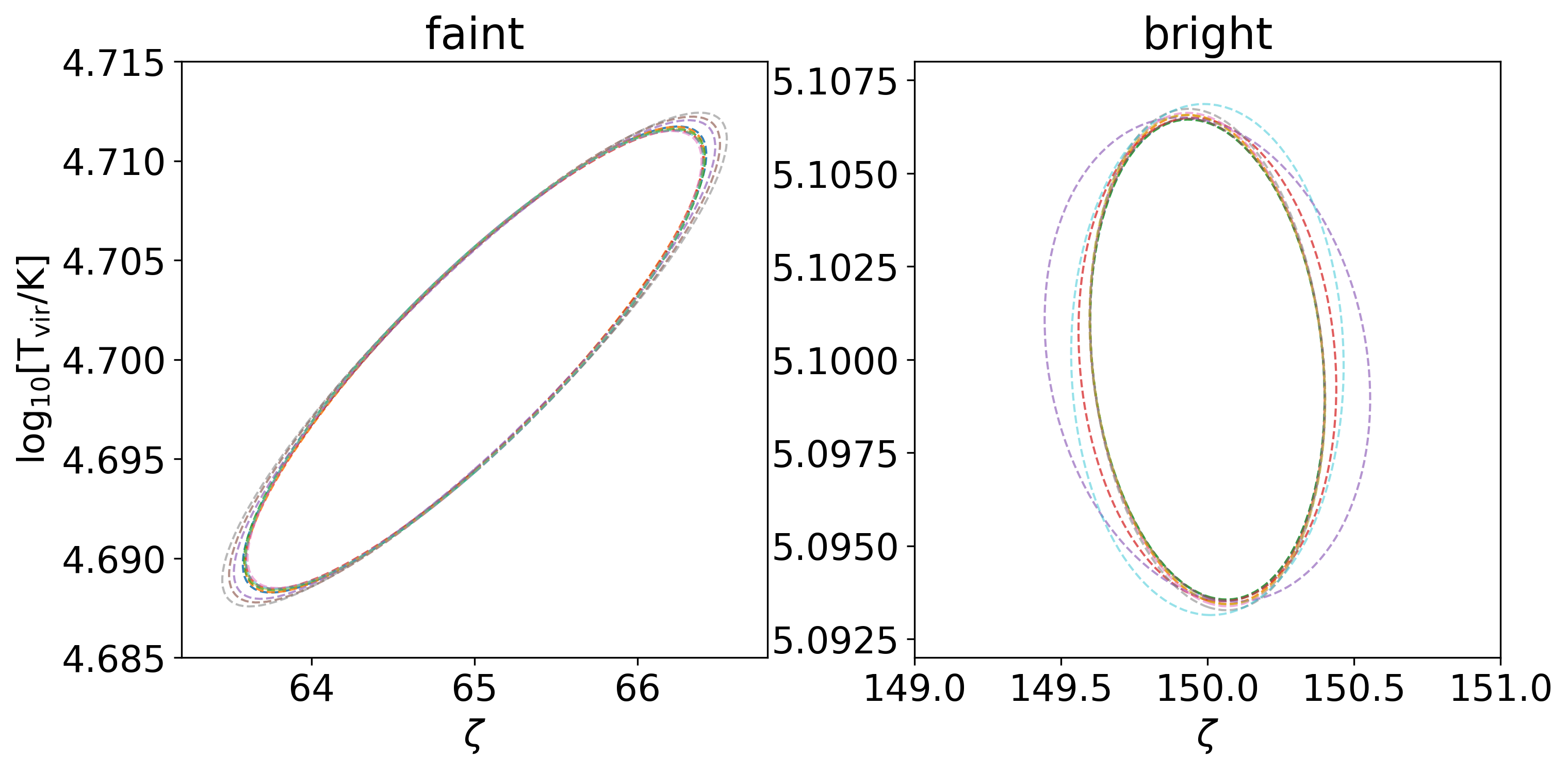}
    \caption{
    The forecasts on the $1\sigma$ confidence region of the parameter constraints when one out of the ten realizations is excluded from the averaging. Each line corresponds to the case where one realization is excluded.
    }
    \label{fig:converge2d}
\end{figure}

 {Furthermore, it is important to test that degeneracy between the parameters is consistent across realizations. To illustrate the convergence of the parameter correlation, we show the covariance of the parameter measurement for $N_{\rm realization} = 9$ in \autoref{fig:converge2d}. As shown, for the faint model, the correlation between the parameters is consistent across all combinations, indicating that the convergence has been reached. The bright model, while also consistent, shows larger variations across different combinations. As we have discussed in \hyperref[sec:values]{Section \ref{sec:values}}, the parameter values are more extreme for the bright model, resulting in larger instabilities. Nevertheless, the parameters are not strongly correlated in the bright model, and the variations do not affect the conclusions reached in this work.}

\section{Forecasts for the foreground wedge at the horizon}
\label{apdx:horizon}
 {As mentioned in \hyperref[sec:discussion]{Section \ref{sec:discussion}}, we can not perform multipole expansion or clustering wedge partition if the foreground wedge is at the physical horizon. For reference, we calculate the Fisher matrix for the spherically averaged monopole for the horizon case, and present the results here.}

\begin{figure}
    \centering
    \includegraphics[width=\linewidth]{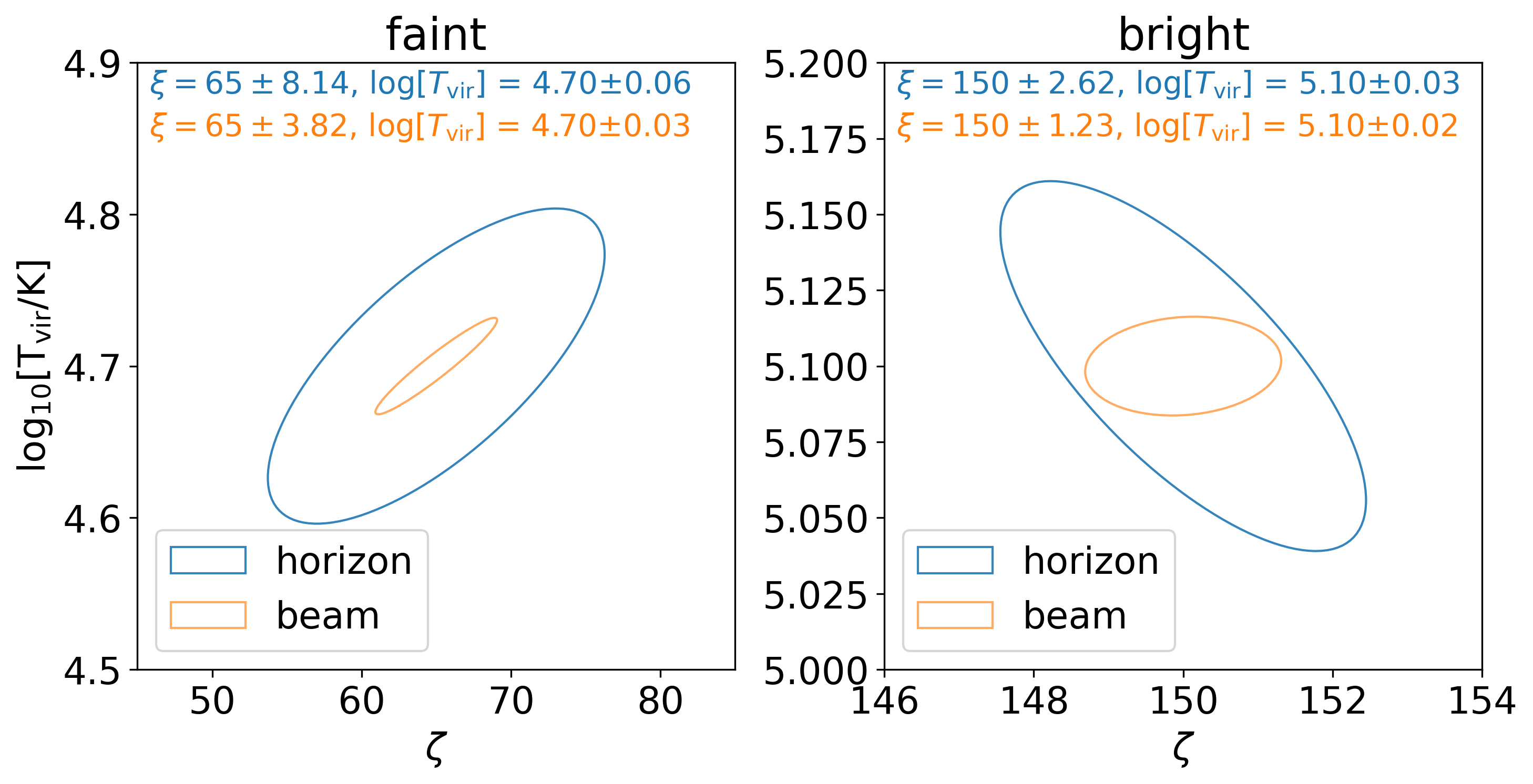}
    \caption{
    The forecasts on the $1\sigma$ confidence region of the parameter constraints for the spherically averaged monopole when the foreground wedge is at the physical horizon. The case where the foreground wedge corresponds to the beam is also presented for comparison. The two cases are colour-coded and the resulting confidence intervals are listed in the top left corner of each panel. 
    }
    \label{fig:horizon}
\end{figure}

 {In \autoref{fig:horizon}, we show the forecasts for the spherically averaged monopole for the horizon case, compared against the case of foreground wedge inside the primary beam field-of-view. For the faint model, the $1\sigma$ confidence interval increases by a factor of $\sim 2.5$. For the bright model, the increase in error bars of the parameters is around a factor of $2$. We note that, since most of the sampling from the baselines centre around $\mu \sim 1$ as shown in \autoref{fig:mu_sample}, the loss in sensitivity in the monopole is not as severe as the $\mu>0.96$ wedge may suggest.}


\bsp	
\label{lastpage}
\end{document}